\documentclass[aps,reprint,twocolumns,pre,superscriptaddress,floatfix,notitlepage,nofootinbib]{revtex4-1}
\usepackage{amssymb,amsmath,amsfonts} 
\usepackage{mathtools}
\usepackage{physics}
\usepackage{array}
\usepackage{graphicx,epsfig,xcolor}
\usepackage[colorlinks=true,citecolor=blue,linkcolor=red, pdftitle = {CorpsRelano-Cat-statesESQPTs}]{hyperref}
\usepackage{newtxtext,newtxmath}
\begin{document}

\newcommand{\hatmath}[1]{\hat{\mathcal{#1}}} 

\title{Energy cat states induced by a parity-breaking excited-state quantum phase transition}

\author{\'{A}ngel L. Corps}
   \email[]{corps.angel.l@gmail.com}
       \affiliation{Instituto de Estructura de la Materia, IEM-CSIC, Serrano 123, E-28006, Madrid, Spain}
       \affiliation{Grupo Interdisciplinar de Sistemas Complejos (GISC), Universidad Complutense de Madrid, Av. Complutense s/n, E-28040 Madrid, Spain}
    
\author{Armando Rela\~{n}o}
\email[]{armando.relano@fis.ucm.es}
\affiliation{Grupo Interdisciplinar de Sistemas Complejos (GISC), Universidad Complutense de Madrid, Av. Complutense s/n, E-28040 Madrid, Spain}
\affiliation{Departamento de Estructura de la Materia, F\'{i}sica T\'{e}rmica y Electr\'{o}nica, Universidad Complutense de Madrid, Av. Complutense s/n, E-28040 Madrid, Spain}

\date{\today} 

\begin{abstract}

We show that excited-state quantum phase transitions (ESQPTs) in a system in which the parity symmetry is broken can be used to engineer an energy cat state ---a Schr\"odinger cat state involving a quantum superposition of both different positions and energies. By means of a generalization of the Rabi model, we show that adding a parity-breaking term annihilates the ground-state quantum phase transition between normal and superradiant phases, and induces the formation of three excited-state phases, all of them identified by means of an observable with two eigenvalues. In one of these phases, level crossings are observed in the thermodynamic limit. These allow us to separate a wavefunction into two parts: one, with lower energy, trapped within one region of the spectrum, and a second one, with higher energy, trapped within another. Finally, we show that a generalized microcanonical ensemble, including two different average energies, is required to properly describe equilibrium states in this situation. Our results illustrate yet another physical consequence of ESQPTs.

\end{abstract}

\maketitle

\section{Introduction}\label{sec:introduction}

One of the most fascinating predictions of quantum mechanics are cat states, named after the famous \textit{Gedankenexperiment} by Schr\"{o}dinger himself \cite{Schrodinger1935}. Cat states are usually defined as macroscopic quantum superpositions of classical states, like particles in different positions. Due to the effects of decoherence,  often induced by a measuring apparatus but also by, e.g., dissipation resulting from the interaction of the quantum system with its environment, these states are very fragile \cite{Wheeler2014}, and thus are not observed classically under normal circumstances\cite{Neumann1932,Zurek1982}. However, they have been generated in the laboratory by means of quantum optics \cite{Ourjoumstev2006,Ourjoumstev2007,Lewenstein2021} or superconducting cavities \cite{Wang2016}.

From the theoretical point of view, a number of techniques to engineer robust cat states have been explored \cite{Pieplow2019,Mateos2021}. A usual one consists in starting in a normal ground state, and then leading the system onto a macroscopic superposition without leaving the ground state, by changing a control parameter \cite{Higbie2004,Huang2006,Nunnenkamp2008,Carr2010}. In many instances, these two kinds of ground states are separated by a quantum phase transition (QPT) \cite{Sachdev}, which is caused by an abrupt, non-analytic change of the ground-state properties of quantum systems, separating two quantum phases characterized by different thermodynamic properties. For example, the ground state of a bosonic Josephson junction made of a number of atoms in a two-site Bose-Hubbard model changes from separable Fock states to Schr\"odinger cat states at a critical value of the interaction amplitude \cite{Mazzarella2011,Relano2014}. This QPT is the basis for the technique proposed in \cite{Huang2006}.

In the last couple of years it has been established that QPTs are not restricted to the ground state. There exists another, perhaps less well-known form of non-analytic behavior detected in the very high-lying (not just slightly above the ground-state) excited states of physical models, giving rise to the phenomenon of excited-state quantum phase transition (ESQPT) \cite{Cejnar2021,Stransky2014,Caprio2008}. ESQPTs have been the subject of intense research during recent years and both their origins and many of their physical consequences have been explored in remarkable detail. These include a number of dynamical effects such as anomalously large decoherence \cite{Relano2008,Perez2009}, singular behavior in quench dynamics \cite{Perez2011,Santos2015,Lobez2016,Bernal2017,Kloc2018}, quantum work
statistics \cite{Wang2017}, and localization \cite{Santos2016}, quantum chaos \cite{Perez2011b,Corps2021arxiv,Bastarrachea2014,Bastarrachea2014II},
the generation of symmetry-breaking equilibrium states
\cite{Puebla2013,Puebla2014}, universal dynamical scaling
\cite{Puebla2020}, dynamical instabilities \cite{Bastidas2014} and dynamical phase transitions \cite{Cabedo2021,Lewis2021},
irreversible processes in which no energy is dissipated \cite{Puebla2015}, and
reversible quantum information spreading \cite{Hummel2019}, to quote a few. For a recent, detailed exposition, we recommend \cite{Cejnar2021}. Nevertheless, some fundamental questions do remain open, among which we highlight the search of a mechanism to link the phenomenology of ESQPTs to that of common QPTs and the definition of truly distinct thermodynamic phases. This is a question that was very recently addressed in Ref. \cite{Corps2021}, on which we will heavily rely in the present work. 

In many cases, QPTs and ESQPTs are closely linked. Let us consider a physical Hamiltonian depending on some control parameter, say $g$, in which a QPT occur at a given critical coupling $g_{c}$. In a large number of collective systems an ESQPT is born after this QPT has been crossed, say $g>g_{c}$, and the corresponding critical energy, $E_c$, merges with the ground state energy at $g=g_c$. The ESQPT is then revealed by non-analyticities in static quantities involving the spectral properties such as the density of states $\varrho(E)$ \cite{Stransky2016,Stransky2015,Macek2019}, but also through equilibrium measurements of relevant physical observables, which may also show a non-analytic behavior at the critical energy. However, this is not necessarily true: it is also possible to find systems with ESQPTs without the corresponding QPT \cite{Relano2016,Stranskyarxiv2021}. 

The main goal of this paper is to take advantage of a similar case to unveil the generation of a macroscopic quantum superposition, occurring both in space and in energy, an energy cat state, by slowly evolving a normal initial state through two ESQPTs of different nature. For this purpose, we focus on a modified version of the Rabi model of quantum optics, for which these ESQPTs can be identified by means of a constant of motion recently proposed in Ref. \cite{Corps2021}. We also provide a statistical ensemble capable of describing the equilibration of physical observables in the long-time dynamics. 

This paper is organized as follows. In Sec. \ref{sec:toymodel} we use a simple classical toy model to illustrate how QPTs can occur in the absence of ESQPTs by studying the critical points of the potential energy. In Sec. \ref{sec:model} we introduce the physical model used in this work, a deformed version of the quantum Rabi model. In Sec. \ref{sec:semiclassical} we analyze the semiclassical features of the quantum model and establish a quantum-classical correspondence. The classical phase space and the level density are studied in Sec. \ref{sec:phasespacedensity}; common indicators of QPTs are used to show that no QPT occurs in the ground-state of the model in Sec. \ref{sec:noqpt}. The features of the quantum model are considered in Sec. \ref{sec:quantumfeatures}, including the expectation value of physical observables in Sec. \ref{sec:expectedvalues}, the level dynamics in Sec. \ref{sec:leveldynamics}, and the analysis of level crossings induced by a parity-breaking ESQPT in Sec. \ref{sec:levelcrossingsfinitesystems}. The generation of energy cat states by the nonequilibrium dynamics and the unitary time evolution is discussed in Sec. \ref{sec:generationcatstates}. We study the thermodynamics of the cat states and we provide a statistical ensemble describing the long-time average of physical observables in these states in Sec. \ref{sec:thermodynamics}. Finally, we gather the main conclusions of our work in Sec. \ref{sec:conclusions}.

\section{Classical toy model}\label{sec:toymodel}
For illustration purposes, let us consider a physical system described by a classical Hamiltonian of the form $H(x,y)=y^{2}/2m+V(x)$ where $m$ is a constant in arbitrary units and $V(x)$ is a \textit{real} analytic function of a single variable, which we call potential. The variables $x$ and $y$ may represent the canonical position and momentum of a classical system of a single degree of freedom, respectively. Following an exposition in line with Ref. \cite{Stransky2014}, suppose that the potential takes the form
\begin{equation}
\label{potential}
    V(x)=x^{4}+bx^{2}+cx,
\end{equation}
where $b,c\in\mathbb{R}$ are some constants. 
All the critical points of these models, including QPTs and ESQPTs, can be found by solving the system of two equations $\nabla H(x_{c},y_{c})=\mathbf{0}$, for $(x_{c},y_{c})$ \cite{Cejnar2021}. In this simple-minded example where $x$ and $y$ are decoupled, one trivially has $y_{c}=0$ and $H(y_{c},x)=V(x)$, so one may focus on the potential $V(x)$ only.  

\begin{center}
\begin{figure}[h!]
\hspace*{-0.60cm}\includegraphics[width=0.54\textwidth]{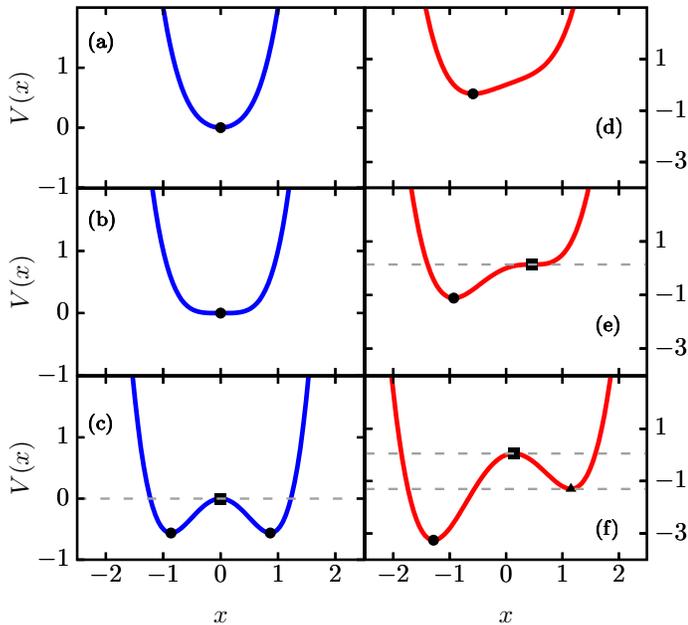}
\caption{Classical potential $V(x)=V(x;b,c)$, Eq. \eqref{potential}, for several values of its parameters. (a-c) The symmetric case $c=0$ for (a) $b=1$, (b) $b=0$ and (c) $b=-3/2$. (d-f) The asymmetric case $c=4/5$ for (d) $b=0$, (e) $b=-1.29266$ and (f) $b=-3$. Black circles mark the critical values $x_{c}$ for which $V(x)$ is minimal (ground-state). Black triangles mark critical points for which $V(x)$ attains other local (not global) minima. Black squares represent critical points for which $V(x)$ attains a local maximum or an inflection point. Triangle and squares can be associated with ESQPTs characterized by a jump discontinuity in the density of states and a logarithmic divergence in the density of states, respectively. Gray dashed lines mark the various energies $V(x)$ for which either kind of ESQPT takes place. }
\label{panelpotential}
\end{figure}
\end{center}

First, let us fix $c=0$ and take $b$ as a control parameter. In this case, $V(x)=V(-x)$, and therefore Eq. \eqref{potential} is a toy-model for one degree of freedom systems with a ${\mathbb Z}_2$ symmetry, like the Lipkin-Meshkov-Glick and the two-fluid Lipkin model, the two-site Bose-Hubbard, the coupled top, the Dicke and the Rabi models \cite{Caprio2008,Relano2008,Perez2009,Kopylov2015,Puebla2013,Perez2017,Feldmann2020,Wang2020,Klocarxiv,Wang2020arxiv,Bastarrachea2014,Brandes2013,Puebla2016,Relano2014,Garcia-Ramos2017,Wang2020}. If $b\geq 0$, the single critical point is $x_{c1}=0$, whereas if $b\leq 0$ there appears a second pair of critical points, $x_{c2,3}=\pm \sqrt{-b/2}$. We therefore identify a critical parameter $b_{c}=0$. Several examples of such a potential for $c=0$ can be seen in Fig. \ref{panelpotential}(a-c). Fig. \ref{panelpotential}(a) shows the case for $b=1$, where there is a single potential minimum. This represents the ground-state of our Hamiltonian. In Fig. \ref{panelpotential}(b) we show $b=b_{c}=0$, which coincides with the value of $b$ for which the second pair of critical points appears. We can see that $V(x)$ shows a single global minimum at $x=0$, much like in the previous panel. However, this critical point is special in that $V(x)$ is completely flat in the neighborhood of $x=0$. The value $b=0$ gives rise to a QPT in the Hamiltonian, by which the ground-state energy shows a non-analytic behavior. We can see in Fig. \ref{panelpotential}(c), for $b=-3/2$, that $V(x)$ admits \textit{two degenerate} global minima (a pairwise degenerate ground-state), while the previous critical point at $x=0$ has turned into a local maximum. This change of behavior occurs exactly at $b=0$, and survives qualitatively for all $b\leq 0$. Therefore, we have the following phase diagram. If $b>0$, we have a unique and symmetric ground state, located at $x=0$. If $b<0$ we have two symmetry-breaking degenerate ground states, and a critical point located at the position of the former ground state, $x=0$, giving rise to an ESQPT. Hence, the critical point, $b_c=0$ accounts for both the QPT and the emergence of the critical ESQPT. It is worth noting that this transition from a symmetric to a symmetry-breaking ground state has been used to build cat states, for example in the two-site Bose Hubbard model \cite{Higbie2004,Huang2006}.

Second, we consider $c>0$, keeping $b$ as a control parameter. Now, $V(x) \neq V(-x)$, and therefore our toy model has no discrete symmetries. The number and value of real critical points in this case depends again on the value of $b$ as $\partial V/\partial x=0$ implies the depressed cubic $x^{3}+v_{1}x+v_{2}=0$ with $v_{1}\equiv b/2$ and $v_{2}\equiv c/4$. The critical coupling $b_{c}$ is the root of the discriminant $\Delta =4v_{1}^{3}+27v_{2}^{2}$, i.e., $b_{c}=-(3/2)c^{2/3}$. For $b>b_{c}$ there is a single critical point $x_{c1}$, while if $b<b_{c}$ there are three different critical points. In Fig. \ref{panelpotential} (d-f) we display this potential for $c=4/5$, for which $b_{c}=-\frac{3 \sqrt[3]{2}}{5^{2/3}}\approx -1.293$. In Fig. \ref{panelpotential}(d) we choose $b=0$, for which the single global minimum at $x_{c1}\approx -0.585$ has potential energy $V(x_{c1})\approx -0.351$. In Fig. \ref{panelpotential}(e), $b=b_{c}$. Apart from the global minimum $V(x_{c1})\approx -1.114 $ at $x_{c1}\approx -0.928$, we find two additional critical points which are equal, $x_{c2,3}\approx 0.464 $ with energy $V(x_{c2,3})\approx 0.139$. Also, the critical points $x_{c2,3}$ are now inflection points. This constitutes the first remarkable consequence of including a symmetry breaking term $c>0$ in Eq. \eqref{potential}. The ground state is always unique and all its properties change smoothly with the control parameter $b$; no traces of a QPT are found. Notwithstanding, $b_c$ still accounts for the emergence of a ESQPT: if $b<b_c$, the potential becomes an asymmetric double well, with a local maximum linked to a critical energy. All these facts impede the creation of a cat state in the ground state, but, as we will see later, open the door to profit from the ESQPT to engineer a different kind of cat state which gives rise to a superposition of different macroscopic energies ---an {\em energy cat} state. Finally, in Fig. \ref{panelpotential}(f), we show the case for $b=-3$. We can find a global minimum $(-1.287, -3.255)$, an additional local minimum $(1.152, -1.298)$, and a saddle point $(0.135, 0.054)$. 

It should be noted that the form of $V(x)$ for $b$ fixed and varying $c$ has been previously studied \cite{Cejnar2008}. In particular, when $b=-1$ and $c$ is taken as a control parameter, Eq. \eqref{potential} does display a \textit{first-order} ground-state QPT at $c_{c}=0$, originating from the swapping of two minima located at $x\neq 0$ within the region defined by $c=\pm \frac{4}{3\sqrt{6}}$. Here we observe, however, that when $c$ is fixed and $b$ varies the situation changes qualitatively.

The results of this introductory section illustrate the consequences of introducing a symmetry-breaking term in a classical potential giving rise to critical phenomena in the ground state and in excited states. In the rest of this work we will focus on a specific quantum system, a modified version of the quantum Rabi model, whose semiclassical analogue shows the same qualitative behavior as this simple potential.

\section{Model: deformed Rabi Hamiltonian}\label{sec:model}

The Rabi model \cite{Rabi1936,Rabi1937} is a paradigmatic system to study both ESQPTs and QPTs. It was originally introduced to describe the interaction between a single bosonic field with frequency $\omega$ and a two-level atom with constant level splitting $\omega_{0}$. We consider a simple generalization of this model introduced in \cite{Corps2021}, which reads
\begin{equation}
\label{eq:model}
\hat{\mathcal{H}}_{\alpha}=\omega\hat{a}^{\dagger}\hat{a}+\omega_{0} \hat{J}_{z}+\sqrt{\omega\omega_{0}}g(\hat{a}^{\dagger}+\hat{a})\hat{J}_{x}+\sqrt{\frac{\omega_{0}}{2}}\alpha(\hat{a}^{\dagger}+\hat{a}),
\end{equation}
where $\hat{a}$ and $\hat{a}^{\dagger}$ are the usual bosonic annihilation and creation operators, $g$ is the coupling strength between the atom and the electromagnetic radiation, and $\mathbf{\hat{J}}$ is the angular momentum for a $j=1/2$ particle. The original Rabi model is recovered when $\alpha=0$. Thus, the last term in the Hamiltonian constitutes a symmetry-breaking deformation, which, as we will see later, entails important qualitative changes in its critical behavior. Other symmetry-breaking deformations have been recently studied, like a term proportional to $\hat{J}_x$ \cite{Braak2011, Liao2016}, or a term proportional to $\left( a^{\dagger} + a \right) \hat{J}_z$
\cite{Stranskyarxiv2021}. We will set $\alpha=1/2$ as a case study. It has been shown \cite{Puebla2016,Hwang2015} that the Rabi model admits a thermodynamic limit, $\omega_{0}/\omega\to\infty$, which coincides exactly with a semiclassical limit. We will fix $\omega=1$, so the TL is reached by simply increasing $\omega_{0}$. For our numerical simulations of the quantum model, Eq. \eqref{eq:model}, we truncate the number of photons to a finite value $n_{ph}$, so the effective Hilbert space dimension is $D=2(n_{ph}+1)$. All numerical results have been tested for convergence and $n_{ph}$ has been optimized. All quantities are expressed in arbitrary units.

Since ESQPTs are known to be deeply rooted in the structure of the semiclassical analogue of the quantum model \cite{Cejnar2021}, we will consider this limit in the next section. We will see that the coupling strength $g$ and the deformation $\alpha$ play a role equivalent to that of $b$ and $c$ in Eq. \eqref{potential}, respectively. 

\section{Semiclassical analysis}\label{sec:semiclassical}

The semiclassical limit of Eq. \eqref{eq:model} is obtained by substituting the photonic operators by the position and momentum operators of the harmonic oscillator, $\hat{p}=i(\hat{a}^{\dagger}-\hat{a})/\sqrt{2}$ and $\hat{q}=(\hat{a}^{\dagger}+\hat{a})/\sqrt{2}$, and then diagonalizing the resulting Hamiltonian matrix \cite{Puebla2016}. On the scale of the reduced energy $\epsilon\equiv E/(\omega_{0}j)=2E/\omega_0$ (with $E$ the actual energy of the system), one obtains the low-energy spin subspace classical Hamiltonian

\begin{equation}
\label{eq:clasicoRM}
    H_{\alpha}(p,q)=\frac{\omega}{\omega_{0}} \left(p^2+q^2 \right) - \sqrt{1+\frac{2\omega g^{2}q^{2}}{\omega_{0}}}+\frac{2 \alpha q}{\sqrt{\omega_{0}}},
\end{equation}
where $(p,q)\in\mathbb{R}^{2}$ are now continuous (non-quantized) classical variables. Thus, the quantum Rabi model Eq. \eqref{eq:model} has a semiclassical analogue of a single effective degree of freedom, $f=1$, and its phase space is $\mathcal{M}=\mathbb{R}^{2}$. Mean-field properties of the quantum model Eq. \eqref{eq:model} such as, e.g., the ground-state energy, the photon population and the atomic population of the ground-state are all appropriately given by the classical analogue Eq. \eqref{eq:clasicoRM} in the limit $\omega_{0}\to\infty$. For finite values of $\omega_{0}$ the quantum model shows corrections with respect to the limiting case of Eq. \eqref{eq:clasicoRM}. 

\subsection{Phase space and density of states}\label{sec:phasespacedensity}

\begin{center}
\begin{figure}[h]
\hspace*{-0.60cm}\includegraphics[width=0.54\textwidth]{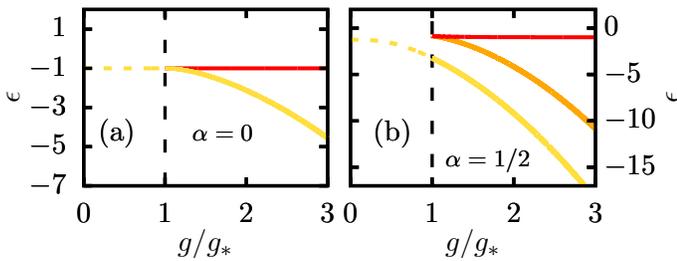}
\caption{Classical energies corresponding to the critical points of the Hamiltonian Eq. \eqref{eq:clasicoRM} as a function of $g/g_{*}$ for (a) the usual Rabi model, $\alpha=0$ and (b) the deformed Rabi model with $\alpha=1/2$. Yellow (light gray) lines represent the ground-state energy, while orange (medium gray) and red (dark gray) lines show the energy at which ESQPTs take place.}
\label{energias}
\end{figure}
\end{center}

Both the ground-state energy, $\epsilon_{\textrm{GS}}$, and the ESQPTs energies $\epsilon_{c1,c2}$ of the system can be obtained as the energies $\epsilon=H_{\alpha}(p_{*},q_{*})$ corresponding to particular \textit{critical points} of Eq. \eqref{eq:clasicoRM} satisfying $\eval{\nabla H_{\alpha}}_{(p_{*},q_{*})}=0$. If $\alpha=0$, these values show an abrupt change at the critical coupling strength $g_{*}(\alpha=0)=1$, which marks a ground-state QPT. If $g\leq g_{*}(\alpha=0)$, the ground-state is $\epsilon_{\textrm{GS}}=-1$, while if $g\geq g_{*}(\alpha=0)$, $\epsilon_{\textrm{GS}}=- (1+g^{4})/2g^{2}$. Besides, if $g\geq g_{*}(\alpha=0)$ there appears a second critical point: it is associated with an ESQPT and corresponds to $\epsilon_{c}=-1$ \cite{Puebla2016,Corps2021}. These results are illustrated in Fig. \ref{energias}(a). This scenario changes qualitatively as soon as $\alpha\neq0$. When $\alpha\neq0$ all critical points and energies can also be obtained analytically but cannot be expressed in terms of elementary functions, so in what follows we will give approximate values to relevant quantities. In Fig. \ref{energias}(b) we show the critical energies for $\alpha=1/2$. We observe that there exists a special coupling strength separating two different regimes: $g_{*}(\alpha=1/2)\approx 1.7872$. For $g<g_{*}(\alpha=1/2)$, there is a single line corresponding to $\epsilon_{\textrm{GS}}$ (which is no longer constant). However, at $g=g_{*}(\alpha=1/2)$ this scenario splits and for $g\geq g_{*}(\alpha=1/2)$ there appear two more energies besides $\epsilon_{\textrm{GS}}$. These two energies grow apart as $g$ increases. This is in contrast with Fig. \ref{energias}(a) where only a single critical excited energy exists.

\begin{center}
\begin{figure}[h]
\hspace*{-0.60cm}\includegraphics[width=0.52\textwidth]{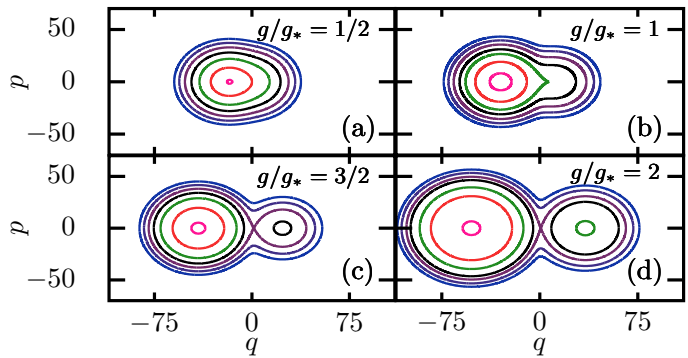}
\caption{Classical phase space of the semiclassical analogue Eq. \eqref{eq:clasicoRM} for $\alpha=1/2$, $\omega=1$, $\omega_{0}=300$ and different values of the coupling strength $g$. Below $g_{*}(\alpha=1/2)\approx 1.7872$ the phase space is compact for all energies, while above $g_{*}$ there is a critical energy whose contour connects two previously disconnected regions.}
\label{contours}
\end{figure}
\end{center}

Both the appearance and the kind of ESQPT a certain critical point in the Hamiltonian flow produces can be understood through the structure of the classical phase space. In Fig. \ref{contours} we show several \textit{classical orbits} of Eq. \eqref{eq:clasicoRM} with $\alpha=1/2$, i.e., the set of points $(p,q)\in\mathbb{R}^{2}$ satisfying $H_{\alpha}(p,q)=\epsilon$. Different lines correspond to different energies. 

In Fig. \ref{contours}(a), we observe that for $g=g_{*}/2<g_{*}$ the classical potential allows for a single, global minimum, corresponding to the ground-state energy. Contour lines simply appear to expand as the energy increases but the structure of the phase space remains unchanged. In contrast to the case $\alpha=0$, the contour curves do not conform a circumference but they are deformed \cite{Corps2021}. This scenario is generic for $g<g_{*}$. This changes dramatically at $g=g_{*}$: Fig. \ref{contours}(b) shows that there exists an energy, which we call \textit{critical}, exhibiting some sort of non-analyticity (a `cusp' can be seen). This energy is associated to an ESQPT (see below). Below this energy, the potential produces a minimum similar to that of Fig. \ref{contours}(a), which is again the ground-state. The remaining panels, Fig. \ref{contours}(c)-(d) concern the case $g>g_{*}$. The structure of the classical phase space is now completely altered: we can see that the potential gives rise to two minima (instead of just one) and a maximum. Such minima are placed at asymmetric values with respect to $q=0$, while the maximum is somewhere near $q=0$. The first minimum is the ground-state energy, $\epsilon_{\textrm{GS}}$, while the second minimum, at $\epsilon_{c1}$, and the single maximum, at $\epsilon_{c2}$, are associated to ESQPTs of different types. We can see that for $\epsilon_{c1}\leq \epsilon\leq \epsilon_{c2}$ the classical phase space is separated into two disconnected regions. The contour line for $\epsilon_{c2}$ crosses itself, giving rise to a singular point in the phase space. A classical trajectory starting from a point in this contour line remains trapped either in the right or in the left well, because the time required to reach the singular point diverges; the same happens regarding the cusp singularity trademark of $g_*$ in Fig. \ref{contours}(b). Above $\epsilon_{c2}$, the phase space acquires compact topology: every two points in an orbit are connected by a contour line\footnote{The phase space is also compact for (i) $g<g_{*}$ at all energies and (ii) for $g>g_{*}$ if $\epsilon_{\textrm{GS}}\leq\epsilon\leq\epsilon_{c1}$. In the first case, the classical potential is of a single-well kind, and in the second case the two-well structure has not been revealed at those low energies.}, and thus every classical trajectory explores both the left and the right parts of the phase space. Therefore, the critical coupling $g_{*}$ marks a transition from single-well to double-well potential, similar to the Rabi model with $\alpha=0$ \cite{Puebla2016,Corps2021}. We can thus see that the behavior of the constant energy curves is qualitatively the same as that shown by the classical cusp potential used as a toy model in Eq. \eqref{potential}.

\begin{center}
\begin{figure}[h]
\hspace*{-0.52cm}\includegraphics[width=0.51\textwidth]{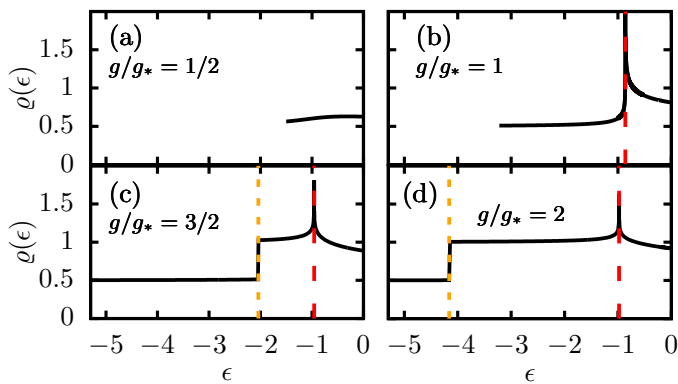}
\caption{Classical level density $\varrho(\epsilon)$ of Eq. \eqref{eq:clasicoRM} for a deformation strength $\alpha=1/2$ and several values of the coupling strength $g$. The finite jump non-analyticity, $\epsilon_{c1}$ is signaled by dotted orange lines, and the logarithmic singularities at $\epsilon_{c2}$, by dashed red lines. These various critical energies are: (b) $\epsilon_{c1}=-0.8620$; (c) $\epsilon_{c1}=-2.0426$, $\epsilon_{c2}=-0.9584$; (d) $\epsilon_{c1}=-4.1596$, $\epsilon_{c2}=-0.9784$. }
\label{density}
\end{figure}
\end{center}

As previously mentioned, the structure of the phase space can be used to ascertain whether ESQPTs exist in the system. In the case of systems with a single degree of freedom, $f=1$, such as Eq. \eqref{eq:clasicoRM}, a characterization of these ESQPTs was presented in \cite{Caprio2008,Cejnar2008} in terms of the various kinds of non-analytic behavior in the classical level density, \begin{equation}\varrho(\epsilon)\equiv \frac{\omega}{\omega_{0}}\frac{1}{2\pi\hbar}\int\textrm{d}p\int\textrm{d}q\,\delta[\epsilon-H_{\alpha}(p,q)].\end{equation}
It is this non-analytic feature of the level density at the ESQPTs critical energies that is most commonly used to diagnose this phenomenon \cite{Cejnar2021}. In our case, the local minima, besides the ground-state, produce \textit{finite jumps} in $\varrho(\epsilon)$, while local maxima give rise to \textit{logarithmic singularities} in $\varrho(\epsilon)$ at the ESQPTs critical energies. To establish a connection with Fig. \ref{contours}, in Fig. \ref{density} we show $\varrho(\epsilon)$ for the same values of the coupling strength $g$ and $\alpha=1/2$. In Fig. \ref{density}(a) we observe a smooth curve without non-analyticities, this scenario being generic as long as $g<g_{*}$. This is because the only fixed point allowed by Eq. \eqref{eq:clasicoRM} when $g<g_{*}$ corresponds to the ground-state energy. In Fig. \ref{density}(b) we show the special case where $g$ coincides exactly with the critical coupling, $g=g_{*}$. We can observe a single, logarithmic singularity in $\varrho(\epsilon)$, signaling an ESQPT. For this coupling strength, the ESQPT critical energies coincide, $\epsilon_{c1}=\epsilon_{c2}$, producing a single singularity in the level density. As exemplified by Fig. \ref{density}(c)-(d), the difference $|\epsilon_{c1}-\epsilon_{c2}|$ grows as $g>g_{*}$ is increased [also see Fig. \ref{energias}(b)]. Indeed, the first critical energy, $\epsilon_{c1}$, corresponds to the second local minima appearing in Fig. \ref{contours}(c)-(d), and it produces a finite jump in $\varrho(\epsilon)$. By contrast, the second critical energy, $\epsilon_{c2}$, corresponds to the local maxima in Fig. \ref{contours}(c)-(d), yielding a logarithmic divergence in $\varrho(\epsilon)$. The various values of the critical energies are indicated in Fig. \ref{density}.

\subsection{Absence of ground-state QPT}\label{sec:noqpt}

Having established the existence of ESQPTs in our model Hamiltonian, we now turn to the following question: Are these ESQPTs connected to a ground-state QPT? The answer, as we will show, is no: the coupling strength $g_{*}$ at which ESQPTs start appearing does not mark any ground-state QPT. What is more: \textit{there is no QPT for any value of $g$} when $\alpha\neq0$. To show this, we consider some common indicators of ground-state QPTs in the Rabi model.

\begin{center}
\begin{figure}[h!]
\hspace*{-0.60cm}\includegraphics[width=0.56\textwidth]{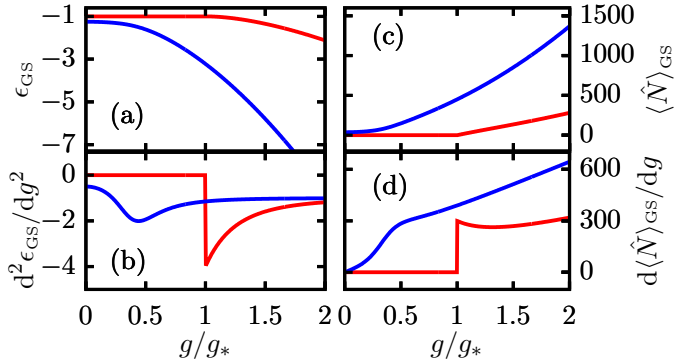}
\caption{Indicators of ground-state quantum phase transitions. (a)-(b): Ground-state energy $\epsilon_{\textrm{GS}}$ and second derivative with respect to the coupling strength $g$. (c)-(d): Number of photons in the ground-state $\langle\hat{N}\rangle_{\textrm{GS}}$ and its derivative with respect to $g$. Red (light gray) curves correspond to the usual Rabi model ($\alpha=0$), while blue (dark gray) curves are for the deformed Rabi model ($\alpha=1/2$). }
\label{noqpt}
\end{figure}
\end{center}

The standard Rabi model (with $\alpha=0$) exhibits a second order ground-state QPT at the critical coupling $g_{*}(\alpha=0)=1$, and this criticality is transferred onto the excited-states in the form of ESQPTs. This QPT is signaled by a non-analyticity in the second derivative of the ground-state energy \cite{Puebla2015}. This can be observed in Fig. \ref{noqpt}(a)-(b), where $\textrm{d}^{2}\epsilon_{\textrm{GS}}/\textrm{d}g^{2}$ becomes discontinuous at $g=g_{*}(\alpha=0)$, even though $\epsilon_{\textrm{GS}}$ appears smooth itself (red [light gray]). Indeed, $\textrm{d}^{2}\epsilon_{\textrm{GS}}/\textrm{d}g^{2}=0$ for $g<g_{*}(\alpha=0)$ while $\textrm{d}^{2}\epsilon_{\textrm{GS}}/\textrm{d}g^{2}=2+3(-1-g^{4})/g^{4}$ for $g>g_{*}(\alpha=0)$, which is indeed discontinuous at $g=g_{*}(\alpha=0)=1$. By contrast, when $\alpha=1/2$ (blue [dark gray]), $\textrm{d}^{2}\epsilon_{\textrm{GS}}/\textrm{d}g^{2}$ is a smooth function of $g$. It is worth mentioning that the discontinuity observed for $\alpha=0$ is smoothed out and transformed into a simple bump observed in Fig. \ref{noqpt}(b).

Another characteristic associated with the ground-state QPT in the Rabi model is the so-called normal-superradiance transition. When $\alpha=0$, for $g<g_{*}(\alpha=0)=1$, the average number of photons in the ground-state of the system, $\hat{N}=\hat{a}^{\dagger}\hat{a}$, is identically 0, $\langle\hat{N}\rangle_{\textrm{GS}}\equiv 0$; this is the normal phase of the model. However, for $g>g_{*}(\alpha=0)$ this number is $\langle\hat{N}\rangle_{\textrm{GS}}>0$ and actually grows boundlessly as a function of $g$ in the superradiant phase. Hence, $\langle\hat{N}\rangle_{\textrm{GS}}$ is a good order parameter, even though it is not linked to the ${\mathbb Z}_2$ symmetry of the model. The average number of photons in the ground-state as well as its derivative with respect to $g$ is represented in Fig. \ref{noqpt}(c)-(d). For $\alpha=0$, we clearly observe that $\langle\hat{N}\rangle_{\textrm{GS}}>0$ is continuous but non-analytic at the critical point; therefore, the QPT is continuous (or second order). Its non-analytic behavior is best seen in Fig. \ref{noqpt}(d), where a finite jump in $\textrm{d}\langle\hat{N}\rangle_{\textrm{GS}}/\textrm{d}g$ is observed. However, when $\alpha=1/2$ we find that $\langle \hat{N}\rangle_{\textrm{GS}}>0$ for \textit{all} values of $g$, i.e., the normal-superradiant phase transition completely disappears. 
Importantly, $\textrm{d}\langle\hat{N}\rangle_{\textrm{GS}}/\textrm{d}g$ also becomes a smooth function when $\alpha=1/2$ for all $g$, leaving no trace of a phase transition whatsoever. 

These results confirm that, in stark contrast with the usual Rabi model ($\alpha=0$), the spectrum of the deformed Rabi model ($\alpha\neq0$) exhibits ESQPTs beyond a coupling strength $g_{*}(\alpha)$ even though there is no QPT in the ground-state for any $g$. That is, its qualitative behavior coincides with that of the toy model discussed in Sec. \ref{sec:toymodel}.

\section{Quantum features of the asymmetric double well structure}\label{sec:quantumfeatures}

The trademark of the deformed version of the Rabi model is the asymmetric double well structure that emerges for $g>g_*$, shown in Fig. \ref{contours}. In this section, we explore its main quantum consequences. As a case study, we again focus on the case $\alpha=1/2$.

\subsection{Expectation values of physical observables}\label{sec:expectedvalues}

\begin{center}
\begin{figure}[h!]
\hspace*{-0.8cm}\includegraphics[width=0.55\textwidth]{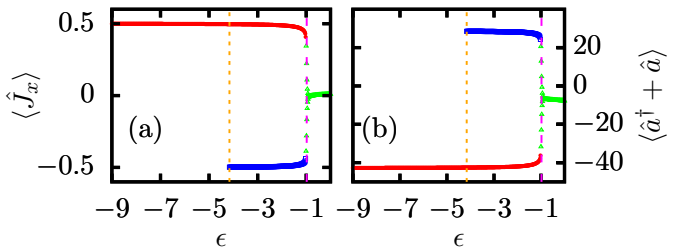}
\caption{Diagonal expectation values of the operators (a) $\hat{J}_{x}$ and (b) $\hat{a}^{\dagger}+\hat{a}$ as a function of energy. Orange (dotted) and magenta (dashed) vertical lines mark the ESQPTs critical energies $\epsilon_{c1}\approx -4.1596$ and $\epsilon_{c2}\approx-0.9785 $, respectively. Model parameters are $\omega=1$ $\omega_{0}=100$, $\alpha=1/2$ and $g/g_{*}=2$. Green (light gray) triangles represent the diagonal expectation values corresponding to eigenstates in the region where the constant $\hat{\mathcal{C}}$ does not exist; red (medium gray) points correspond to eigenstates for which $\bra{\epsilon}\hat{\mathcal{C}}\ket{\epsilon}\leq -0.95$; and blue (dark gray) points to eigenstates for which $\bra{\epsilon}\hat{\mathcal{C}}\ket{\epsilon}\geq 0.95$. Number of photons is $2000$. }
\label{fig:obs}
\end{figure}
\end{center}

A first idea of the consequences of the asymmetric double well structure can be obtained by studying the diagonal expectation values of representative observables in the eigenstates of the Hamiltonian,
$\langle\hat{\mathcal{O}_n}\rangle\equiv\bra{\epsilon_{n}}\hat{\mathcal{O}}\ket{\epsilon_{n}}$. We have worked with $g/g_*=2$, and $\omega_0=100$, and we have chosen $\hat{J}_{x}$ and $\hat{a}^{\dagger}+\hat{a}$ as representative observables. Results are shown in Fig. \ref{fig:obs}. We observe a first remarkable outcome. For energies $\epsilon_{c1} \leq \epsilon \leq \epsilon_{c2}$, both observables show a double-branch structure. It is enlightening to compare this result with the contour plots shown for the classical phase space in Fig. \ref{contours}(d), corresponding to the same value of $g$. Between $\epsilon_{c1}$ and $\epsilon_{c2}$, classical trajectories are trapped either on the left ($q<q_c$) or in the right ($q>q_c$) part of the phase space, and below $\epsilon_{c1}$ all the trajectories are trapped on the left part. This is exactly what happens with the expectation value of $\hat{a}^{\dagger} + \hat{a}$, which is proportional to the operator $\hat{q}$. Therefore, results in Fig. \ref{fig:obs} suggest that there is a direct link between the topology of the classical trajectories and the properties of quantum eigenstates. An important consequence of these results is their incompatibility with the eigenstate thermalization hypothesis (ETH) \cite{Alessio2016,Tasaki1998,Rigol2008,Reimann2015,Deutsch2018,Srednicki1994}. In particular, for the long-time average of physical observables to coincide with a suitable microcanonical average around a target energy, the ETH requires the diagonal matrix elements of the observables to vary only smoothly with energy. However, in our case, Fig. \ref{fig:obs} shows abrupt variations in the diagonal matrix elements when $\epsilon_{c1}\leq \epsilon\leq \epsilon_{c2}$.   Therefore, neither the microcanonical, nor the standard Gibbs ensemble are expected to hold for this system, even though it has neither discrete symmetries, nor other operators commuting with the Hamiltonian. 

This idea has been exploited in \cite{Corps2021} to propose that a large class of ESQPTs can be identified by means of a constant of motion holding just below the corresponding critical energy. Due to the properties of the classical trajectories, the operator
\begin{equation}\label{constant}
    \hat{\mathcal{C}}\equiv \textrm{sign}(\hat{q}-q_{c}\mathbb{I})
\end{equation}
is proposed as a constant of motion below $\epsilon_{c2}$ and in the thermodynamic limit. Here, $q_{c}$ is the classical position corresponding to the critical energy $\epsilon_{c2}$ and $\mathbb{I}$ is the identity matrix. Details on the computation of Eq. \eqref{constant} can be found in Refs. \cite{Corps2021,Corps2021arxiv}. If $\epsilon_n < \epsilon_{c2}$, then $\left[\left| \epsilon_n \right> \left< \epsilon_n \right|, \hat{{\mathcal C}} \right]=0$, where $\hat{\mathcal{H}}\ket{\epsilon_{n}}=(\omega_{0}/2)\epsilon_{n}\ket{\epsilon_{n}}$. As $\hat{{\mathcal C}}$ has just two eigenvalues, $\textrm{Spec}\, (\hat{{\mathcal C}})=\pm 1$, this means that $\bra{\epsilon_{n}}\hatmath{C}\ket{\epsilon_{n}}=-1$ if the eigenstate $\left| \epsilon_n \right>$ is attached to the left part of the classical phase space ($q<q_c$), and $\bra{\epsilon_{n}}\hatmath{C}\ket{\epsilon_{n}}=1$, if it is attached to the right part ($q>q_c$). We have used this fact to choose the color points in Fig. \ref{fig:obs}: red (medium gray) points represent eigenstates with $\bra{\epsilon_{n}}\hatmath{C}\ket{\epsilon_{n}}=-1$; blue (dark gray) points, eigenstates with $\bra{\epsilon_{n}}\hatmath{C}\ket{\epsilon_{n}}=1$, and green (light gray) triangles, eigenstates above $\epsilon_{c2}$, for which $\hat{{\mathcal C}}$ is no longer a constant of motion. We can see that this theory provides a perfect explanation for the structure displayed in Fig. \ref{fig:obs}. Therefore, we can gather the following important conclusion. {\em Below $\epsilon_{c2}$, we have two independent sets of eigenstates (at least in the thermodynamic limit): one characterized by $\bra{\epsilon_{n}}\hatmath{C}\ket{\epsilon_{n}}-1$, and another one characterized by $\bra{\epsilon_{n}}\hatmath{C}\ket{\epsilon_{n}}=1$.} That is, we have an extra quantum number, $\pm 1$, to label all the eigenstates below $\epsilon_{c2}$. As we will see later, this observation is of capital importance for dynamics across the ESQPT.

\subsection{Level dynamics}\label{sec:leveldynamics}

\begin{center}
\begin{figure}[h!]
\hspace*{-0.60cm}\includegraphics[width=0.54\textwidth]{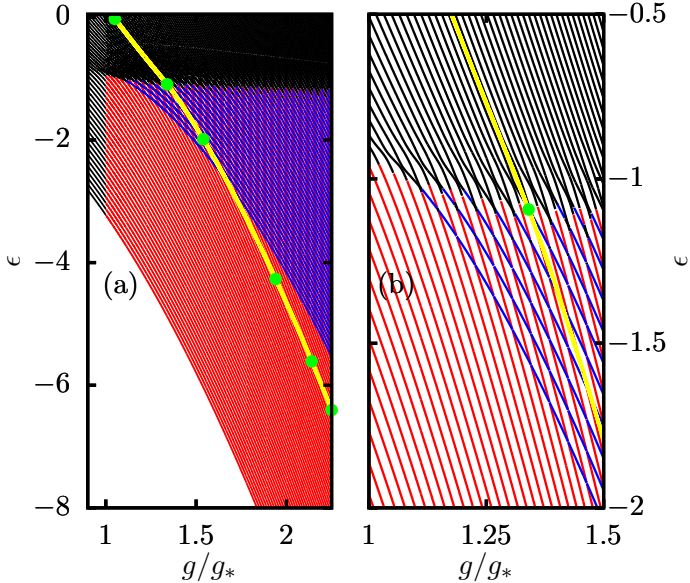}
\caption{(a) Energy flow diagram of the quantum model Eq. \eqref{eq:model} as a function of the coupling strength, $\epsilon=\epsilon(g/g_{*})$. The lowest energy line represents the ground-state, while upper lines show increasingly high excited states of the system. Model parameters are $\omega=1$, $\omega_{0}=20$, and $\alpha=1/2$. Black lines represent energy levels in the region where the constant $\hat{\mathcal{C}}$ does not exist; red (light gray) lines show energy levels for which $\bra{\epsilon}\hat{\mathcal{C}}\ket{\epsilon}\leq -0.95$; and blue (dark gray) lines the energy levels for which $\bra{\epsilon}\hat{\mathcal{C}}\ket{\epsilon}\geq 0.95$. The yellow thick line shows the energy of the quenched state $\epsilon_{g}=\bra{\Psi(g)}\hat{\mathcal{H}}(g)\ket{\Psi(g)}$. Green points show special cases $(g/g_{*},\epsilon_{g})$ whose energy distribution is shown in Fig. \ref{fig:distributions}. (b) Magnification of the flow diagram. Number of photons is $570$.  }
\label{fig:flow}
\end{figure}
\end{center}

The next step consists in studying the dynamical consequences of the previous results. We start by considering the level flow diagram of the quantum model Eq. \eqref{eq:model} shown in Fig. \ref{fig:flow}. Such a diagram displays the value of several energy levels $\epsilon_{n}$ ($n=1,2,...$) as a function of the coupling strength $g$, and it gives relevant insight into the dynamics of energy levels. The first energy line represents the ground-state energy of $\hat{\mathcal{H}}(g)$ for each $g$, and higher energy excited-states follow upwards. We observe that the structure of the level flow closely resembles the classical picture in Fig. \ref{energias}(b). Namely, energy levels show a collapse onto a single line around $\epsilon_{c2}\approx-1$ (though this value is not exactly constant for different $g$) which is a clear signature of a logarithmic divergence of the level density $\varrho(\epsilon)$ in an ESQPT [see Fig. \ref{density}]. We can also see in the magnified picture of Fig. \ref{fig:flow}(b) that apparent level crossings happen in an energy range below $\epsilon_{c2}$. The energy below which this phenomenon stops happening, $\epsilon_{c1}$, marks another ESQPT signaled by a finite discontinuity in the level density [Fig. \ref{density}]. 

As in Fig. \ref{fig:obs}, we have used the constant of motion $\hat{{\mathcal C}}$ to characterize the energy levels in the flow diagram of Fig. \ref{fig:flow}. Black lines show the energy levels where the quantum operator Eq. \eqref{constant} is not a constant of motion: this happens (i) for $g<g_{*}$ at all energies, and (ii)  for $g>g_{*}$ only at energies $\epsilon>\epsilon_{c2}$. Since the ratio $\omega_{0}=20$ is far from the thermodynamic limit $\omega_{0}\to\infty$, some finite-size effects exist. Red (light gray) lines show energy levels whose eigenstates belong to the left classical well, $\langle\hat{\mathcal{C}}\rangle=-1$, while blue (dark gray) lines show levels whose eigenstates belong to the right classical well, $\langle \hat{\mathcal{C}}\rangle=+1$. Below $\epsilon_{c1}$ all levels belong to the left well because the right well only appears at $\epsilon_{c1}$, but between $\epsilon_{c1}$ and $\epsilon_{c2}$ they can belong to either well, as the diagram clearly shows. These two classes of levels are the ones which \textit{appear to cross} within the region $\epsilon_{c1}\leq\epsilon\
\leq\epsilon_{c2}$. 

This would be compatible with the existence of two independent sets of eigenstates, as we have proposed at the end of the previous section. However, the Neumann-Wigner theorem \cite{Neumann1929,Landau1965} states that exact crossings are possible only if there exists an exact quantum number labeling the crossing levels unambiguously. For example, in \cite{Nader2021} level crossings occur between states of different parity of the Lipkin-Meshkov-Glick model, while levels belonging in the same parity subspace produce anticrossings. As results in \cite{Corps2021} indicate that $\hat{{\mathcal C}}$ becomes an exact constant of motion only in the thermodynamic limit, it is reasonable to expect that apparent crossings in Fig. \ref{fig:flow} are not exact, but avoided. To shed some light on this important issue, and to get an idea of the expected finite-size effects, we rely on the semiclassical approximation. As shown in \cite{RelanoEPL,Bastarrachea2017}, we can use the standard Einstein-Brillouin-Keller (EBK) action quantization rules \cite{EBK} to determine the positions of the energy levels of Eq. \eqref{eq:model} in the thermodynamic limit. To do so, we rely on the semiclassical model, Eq. \eqref{eq:clasicoRM}, to solve the following integral equation
\begin{equation}
\label{eq:requ}
\oint_{\Gamma_{\pm}} \, p(\epsilon,q) \textrm{d}q = 2 \pi (n + a_n),    
\end{equation}
where $p(\epsilon,q)$ is obtained by inverting Eq. \eqref{eq:clasicoRM} for a given energy $\epsilon$. The values $\epsilon_n$ for which $n \in {\mathbb N}$ provide the energy levels of our model in the thermodynamic limit. The integral is performed over a closed trajectory $\Gamma_{\pm}$ covering either the left ($-$) or the right ($+$) part of the phase space. The quantities $a_n$ are related to the Maslov index of the corresponding trajectory, which are equal to $1/4$ for each turning point; thus, $a_n=1/2$, $\forall n$. Since the vacuum energy of the harmonic oscillator, $E=\omega/2$, is removed from the Rabi model, the energy obtained from Eq. \eqref{eq:requ} has to be shifted accordingly.

\begin{center}
\begin{figure}[h!]
\hspace*{-0.60cm}\includegraphics[width=0.54\textwidth]{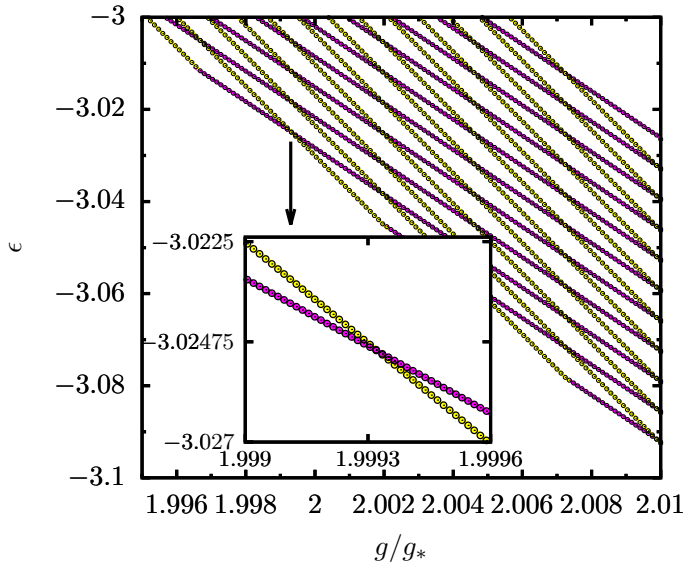}
\caption{Region of level flow diagram for $1.995\leq g \leq 2.01$. Model parameters are $\omega=1$, $\omega_{0}=300$, and $\alpha=1/2$. Numerical energy levels, corresponding to the eigenvalues of the quantum model Eq. \eqref{eq:model}, are shown with empty circles. Color lines represent the classical energy levels obtained via quantization, Eq. \eqref{eq:requ}; yellow (light gray) lines correspond to levels with $\langle \hat{\mathcal{C}}\rangle=-1$, and magenta (dark gray) lines to $\langle\hat{\mathcal{C}}\rangle=+1$. Inset: magnification of a single level crossing. The number of photons goes up to $5000$.  }
\label{fig:requantized}
\end{figure}
\end{center}

In Fig. \ref{fig:requantized} we compare the theoretical prediction given by Eq. \eqref{eq:requ} for $\omega_0=300$ and $1.995 \leq g/g_* \leq 2.010$, with the eigenvalues obtained by diagonalizing Eq. \eqref{eq:model}. The theoretical, semiclassical curves undergo several exact crossings, since the integrals for $\langle \hat{{\mathcal C}} \rangle= -1$ and $\langle \hat{{\mathcal C}} \rangle= 1$ arise from different closed paths $\Gamma_{\pm}$, and therefore they are independent. We can see that the numerical energy levels are almost indistinguishable from the corresponding theoretical predictions for this range of parameters. In particular, we display in the inset a single level crossing, with $1.9990 \leq g/g_* \leq 1.9996$; the differences between the EBK and the exact energy levels are negligible altogether.

To interpret the consequences of this result, we come back to Fig. \ref{fig:flow}. We can see there that the rapidity at which the value of every energy level changes with $g$, $\epsilon_n(g)$, clearly depends on $\bra{\epsilon_{n}}\hatmath{C}\ket{\epsilon_{n}}$; in particular, the `speed' $\left| \textrm{d} \epsilon_{n,-}/\textrm{d}g\right|>\left| \textrm{d} \epsilon_{n,+}/\textrm{d}g\right|$, where the subindex $-$ identifies the energy levels with $\bra{\epsilon_{n}}\hatmath{C}\ket{\epsilon_{n}}=-1$, and the subindex $+$, the energy levels with $\bra{\epsilon_{n}}\hatmath{C}\ket{\epsilon_{n}}=+1$. This fact, together our previous statement concerning the almost exactitude of the EBK rules for not so large values of $\omega_0$, motivate us to formulate the following conjecture:

{\em Even in a finite-size system, the energy levels with $\langle \hat{{\mathcal C}} \rangle=-1$ evolve independently of the energy levels with $\langle \hat{{\mathcal C}} \rangle=1$, when changing the value of the coupling constant $g$. Therefore, we expect the following for an adiabatic evolution. If the wavefunction consists in a superposition of states with both $\langle \hat{{\mathcal C}} \rangle=\pm 1$, the energy  of  the states with $\langle \hat{{\mathcal C}} \rangle=-1$ will change faster with $g$, and therefore an adiabatic passage across the region with $\epsilon_{c1} \leq \epsilon \leq \epsilon_{c2}$ can produce a superposition of different macroscopic energies.}

In Sec. \ref{sec:levelcrossingsfinitesystems} we will provide evidence in favor of this intuition, including a finite-size scaling of the behavior of the level crossings. 

\subsection{Level crossings in finite-size systems}\label{sec:levelcrossingsfinitesystems}

In Sec. \ref{sec:generationcatstates} we will use the previous results to generate an energy cat by slowly passing through the ESQPTs. But before that, it is interesting to investigate the behavior of two consecutive energy levels when crossing the critical energy in a finite-size system, in order to foresee the possible consequences of such an adiabatic passage. 

\begin{center}
\begin{figure}[h!]
\hspace*{-1cm}\includegraphics[width=0.58\textwidth]{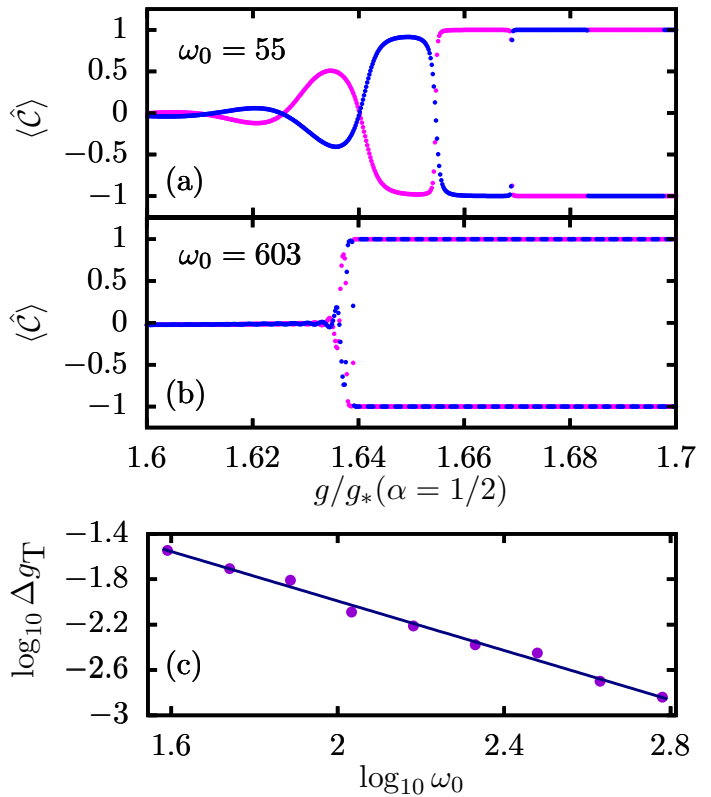}
\caption{(a-b) Diagonal expectation values of $\hat{\mathcal{C}}$ in the eigenstate $\ket{\epsilon_{p}(g)}$ (blue [dark gray]) such that $\epsilon_{p}(g/g_{*}=1.6)\approx -0.8$ and in the eigenstate with energy closest and below the previous level, $\ket{\epsilon_{p-1}(g)}$ (magenta [light gray]), as the coupling parameter $g$ is varied. The system-size parameter leading to the thermodynamic limit is $\omega_{0}=55$ in (a) and $\omega_{0}=603$ in (b). (c) Finite-size scaling of the amplitude of the transient behavior region $\Delta g_{\textrm{T}}$. We obtain the power-law behavior $\Delta g_{\textrm{T}}\sim 1/\omega_{0}^{z}$ with $z\approx 1$. The number of photons varies in $n_{ph}\in[812,8977]$ depending on $\omega_0$. }
\label{fig:panelcrucescnn}
\end{figure}
\end{center}

We first focus on how the diagonal expectation values of $\hat{\mathcal{C}}$ change for a given pair of energy levels as the coupling $g$ is varied. We consider a narrow coupling strength span $g/g_{*}(\alpha)\in[1.6,1.7]$. For each value of the system-size parameter $\omega_{0}$, we look for the energy level $\epsilon_{p}$ closest to energy $\epsilon=-0.8$ at $g/g_{*}(\alpha)=1.6$, whose position in the energy spectrum is denoted by $p\in\mathbb{N}$.
Then we consider the diagonal expectation value of $\hat{\mathcal{C}}$ in two eigenstates for different values of $g$: the eigenstate with energy $\epsilon_{p}$, denoted $\ket{\epsilon_{p}(g)}$ for each value of the coupling $g$, and also that  corresponding to $\epsilon_{p-1}$, the eigenlevel closest and below the previous energy level,  $\ket{\epsilon_{p-1}(g)}$. That is, we focus on $\bra{\epsilon_{p}(g)}\hat{\mathcal{C}}\ket{\epsilon_{p}(g)}\equiv \langle \hat{\mathcal{C}}\rangle_{p,g}$ and $\bra{\epsilon_{p-1}(g)}\hat{\mathcal{C}}\ket{\epsilon_{p-1}(g)}\equiv \langle\hat{\mathcal{C}}\rangle_{p-1,g}$. In other words, we choose two given energy levels at an initial value of the coupling parameter and then follow the evolution of those same levels as the coupling is increased. The curvature of the level flow diagram shown in Fig. \ref{fig:flow} implies that as $g$ increases the two energy levels $\epsilon_{p}(g)$ and $\epsilon_{p-1}(g)$, which are above $\epsilon_{c2}$ for $g/g_{*}(\alpha)=1.6$, will eventually cross the ESQPT at energy $\epsilon_{c2}$ for some $g$. Once the ESQPT has been crossed, $\hat{\mathcal{C}}$ acts as a very approximate constant of motion, though it should be emphasized that the exact constancy of $\hat{\mathcal{C}}$ only occurs in the thermodynamic limit \cite{Corps2021}. After the ESQPT at energy $\epsilon_{c2}$ has been crossed, the (diagonal) expectation values of $\hat{\mathcal{C}}$ in the eigenstates of $\hat{\mathcal{H}}_{\alpha}(g)$ considered can only be $-1$ and $+1$, as explained before. These expectation values are depicted in Fig. \ref{fig:panelcrucescnn}(a)-(b) for $\omega_{0}=55$ (a) and $\omega_{0}=603$ (b). Let us focus on panel (a). We observe that for sufficiently large $g$, $\langle \hat{\mathcal{C}}\rangle_{p,g}$ changes abruptly between $-1$ and $+1$ at given values of $g$, and the same is true for $\langle \hat{\mathcal{C}}\rangle_{p-1,g}$. This means that at those values of $g$, there occurs a precursor of a level crossing which induces a swapping of these conserved quantities. Below a certain $g$, these expectation values are not simply $-1$ and $+1$. The reason is that for those $g$ the operator $\hat{\mathcal{C}}$ is still not a constant of motion at the considered energies $\epsilon_{p}(g)$ and $\epsilon_{p-1}(g)$. This is clear for the smaller values of $g$, since $\epsilon_{p}(g),\epsilon_{p-1}(g)\approx -0.8$ are significantly above $\epsilon_{c2}$. Moreover, there is a more interesting intermediate region of $g$ where $\hat{{\mathcal C}}$ shows a transient behavior. This is because the ESQPT, as any other phase transition, only truly happens in the thermodynamic limit, so its effects can be blurred in the quantum model even when the ESQPT critical energy of the thermodynamic limit, $\epsilon_{c2}$, has been crossed. To perform a quantitative analysis, we define the $g$-width $\Delta g_{T}\equiv g_{0.05}-g_{0.95}$, where $g_{0.05}$ stands for the last value of the coupling $g$ such that $1-\langle \hat{\mathcal{C}}\rangle_{p,g}^{2}\geq 0.05$, and $g_{0.95}$ represents the first $g$ such that $1-\langle \hat{\mathcal{C}}\rangle_{p,g}^{2}\leq  0.95$. Then, we study how this magnitude changes with $\omega_0$ by displaying $\Delta g_{\textrm{T}}$ as a function of $\omega_{0}$ in Fig. \ref{fig:panelcrucescnn}(c). We obtain the behavior $\Delta g_{\textrm{T}}\sim 1/\omega_{0}^{z}$ with $z\approx 1$, strongly suggesting the shrinking $\Delta g_{\textrm{T}}\to 0$ as a power-law approaching the thermodynamic limit $\omega_{0}\to\infty$. This is fully compatible with the results published in \cite{Corps2021}. As the thermodynamic limit is approached, the change from a region in which the value of $\hat{{\mathcal C}}$ in an eigenstate of the Hamiltonian is totally undefined, and $\langle \hatmath{C} \rangle^{2} - \langle \hatmath{C}^2 \rangle = 1-\langle \hatmath{C}^2 \rangle \sim 1$, to a region in which $\hatmath{C}$ is perfectly defined, and $\langle \hatmath{C} \rangle^{2} - \langle \hatmath{C}^2 \rangle = 1-\langle \hatmath{C}^2 \rangle \sim 0$, becomes more abrupt. Afterwards, the number of (almost exact) level crossings increase with $\omega_0$.

\begin{center}
\begin{figure}[h!]
\hspace*{-1.3cm}\includegraphics[width=0.56\textwidth]{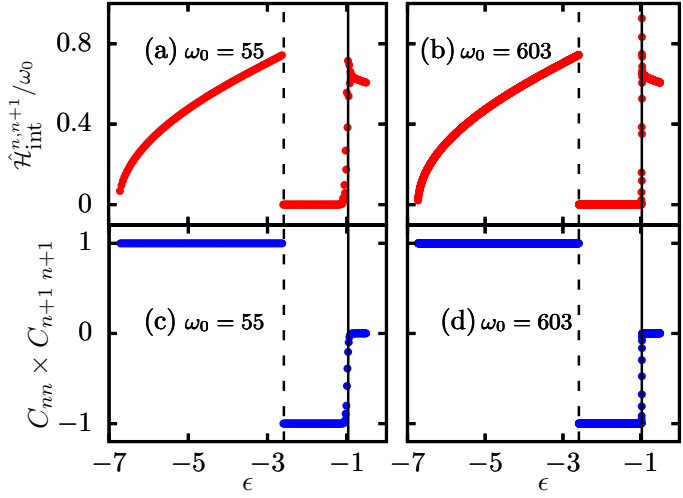}
\caption{(a-b) Expectation values of the interaction term of the Hamiltonian $\hat{\mathcal{H}}_{\textrm{int}}$ in adjacent energy eigenstates of the full Hamiltonian Eq. \eqref{eq:termsadiabatic} as a function of energy for (a) $\omega_{0}=55$ and $\omega_{0}=603$. (c-d) Product of the diagonal expectation values of $\hat{\mathcal{C}}$ in adjacent energy eigenstates, $C_{nn}$ and $C_{n+1\,n+1}$, as a function of energy, for (c) $\omega_{0}=55$ and (d) $\omega_{0}=603$. Model parameters are $\omega=1$, $\alpha=1/2$, and $g/g_{*}=1.65$. Black dashed lines mark the energy of the first ESQPT, $\epsilon_{c1}=-2.5886$, while full black lines mark the second ESQPT, $\epsilon_{c2}=-0.9668$. Number of photons ranges in $n_{ph}\in[752,8135]$ depending on $\omega_0$. }
\label{fig:paneladiabatic}
\end{figure}
\end{center}

Let us now suppose that we start in an eigenstate of the Hamiltonian at a given value of the coupling constant $g$, with an energy above $\epsilon_{c2}$, and that we perform a time-dependent protocol $g(t)$ with the aim of crossing the ESQPT of energy $\epsilon_{c2}$. We can write the time-dependent wavefunction $\left| \Psi(t) \right> = \sum_n c_n (t) \left| \epsilon_n (t) \right>$, where $\left| \epsilon_n (t) \right>$ are the instantaneous eigenstates for Eq. \eqref{eq:model} with $g(t)$. Then, the coefficients $c_n(t)$ evolve according to 
\begin{equation}
\label{eq:adi}
\begin{split}
&\dot{c}_m(t) + \left[ i \epsilon_m(t) + \left<\epsilon_m(t) \right| \left. \dot{\epsilon_m}(t) \right> \right]c_m(t) = \\ &= \sum_{n \neq m} \frac{\left< \epsilon_m(t) \right| \dot{\mathcal{H}} \left| \epsilon_n(t) \right>}{\epsilon_m(t) - \epsilon_n(t)} c_n(t),
\end{split}
\end{equation}
where the overdot indicates a time derivative, and we have set $\hbar\equiv 1$. The first line in Eq. \eqref{eq:adi} accounts for the phase acquired by any coefficient $c_n(t)$ as a result of the time evolution, whereas the second line accounts for the non-adiabatic transitions between the instantaneous energy levels. Therefore, to estimate the relevance of such transitions in finite-size systems, we study the following magnitude,
\begin{equation}\label{eq:termsadiabatic}
   \hat{\mathcal{H}}_{\textrm{int}}^{n,n+1}\equiv  \frac{\bra{E_{n+1}(g)}\hat{\mathcal{H}}_{\textrm{int}}\ket{E_{n}(g)}}{E_{n}(g)-E_{n+1}(g)}.
\end{equation}
with
\begin{equation}
\label{eq:interaction}
    \hat{\mathcal{H}}_{\textrm{int}}=\sqrt{\omega\omega_{0}}(\hat{a}^{\dagger}+\hat{a})\hat{J}_{x},
\end{equation}
that it is proportional to the contribution of the neighboring energy levels to the non-adiabatic transitions in time-dependent protocols $g(t)$. We represent this magnitude in Fig. \ref{fig:paneladiabatic}(a-b) for two values of the thermodynamic limit parameter, $\omega_{0}=55$ and $603$, and $g/g_*=1.65$. Moreover, in Fig. \ref{fig:paneladiabatic}(c-d) we represent the product of the diagonal expectation value $C_{nn}=\bra{E_{n}}\hat{\mathcal{C}}\ket{E_{n}}$ in adjacent eigenstates, for the same values of $\omega_{0}$. Both cases show the same qualitative result. We observe that for $\epsilon_{\textrm{GS}}\leq \epsilon\leq \epsilon_{c1}$ the expectation values are non-vanishing, and thus transitions are possible between the states $E_{n}$ and $E_{n+1}$. Within this energy window all eigenstates belong to the same symmetry subspace, in particular they all have $\langle \hat{\mathcal{C}}\rangle=-1$ and therefore $C_{nn}\times C_{n+1\,n+1}=+1$ as shown in the lower panels. By contrast, in the region $\epsilon_{c1}\leq \epsilon\leq \epsilon_{c2}$ between both ESQPTs we observe that the transition amplitudes vanish, indicating that there are no allowed transitions between adjacent eigenstates in this region. This is because in this energy region adjacent eigenstates effectively belong to different symmetry subspaces, as they are very approximately characterized by $\langle \hat{\mathcal{C}}\rangle$ of opposite sign, $\pm 1$. This is displayed in the lower panels, which show that for adjacent eigenstates $C_{nn}\times C_{n+1\,n+1}=-1$. For $\epsilon\geq \epsilon_{c2}$, $\hat{\mathcal{C}}$ no longer acts as a constant of motion and therefore the classification of eigenstates in symmetry sectors no longer holds. We observe that when $\epsilon\geq \epsilon_{c2}$ transitions between adjacent eigenstates are again allowed. 

\begin{center}
\begin{figure}[h!]
\hspace*{-0.60cm}\includegraphics[width=0.5\textwidth]{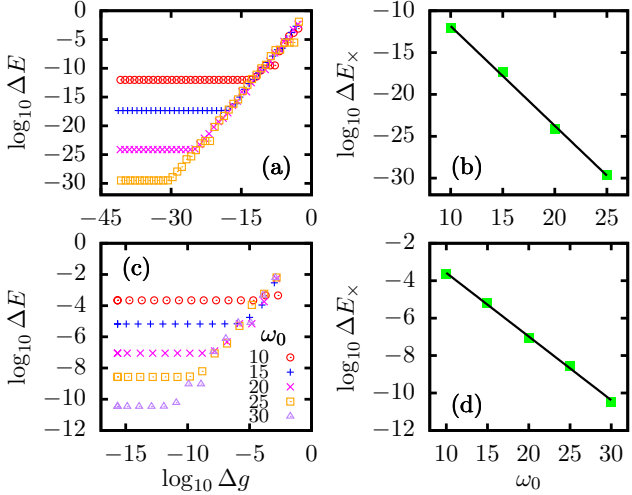}
\caption{(a,c) Estimation of the minimum distance between levels at an avoided crossing, $\Delta E$, as a function of the control parameter width, $\Delta g$ for $\alpha=1/2$. Results for different values of the thermodynamic limit parameter $\omega_{0}$ (see legend) are shown with different symbols. The crossings are all located around $\epsilon\approx -4$ in (a), while in (c) they are at $\epsilon\approx -2$. (b,d) Scaling of the saturation of the minimum distance at an avoided crossing as a function of $\omega_{0}$ (squares) for the crossings in (a,c), respectively. In (b), the black line represents the exponential decay $\Delta E_{\times}(\omega_{0})\sim 10^{-\delta \omega_{0}}$ with $\delta\approx 1.19$, while in (d) the exponent is $\delta\approx 0.34$. }
\label{fig:deltaEcrossings}
\end{figure}
\end{center}

Before ending this section, we provide an additional exploration of the behavior of level crossings in finite-$\omega_{0}$ systems. Now, instead of studying the behavior of an energy level through several level crossings, we will focus on a single one of these crossings. Assuming that a Landau-Zener transition \cite{LandauZener,Landau1965} consecutively mixes states with different quantum numbers, $\langle\hatmath{C}\rangle$, the probability of a non-adiabatic transition around a typical crossing can be estimated to be $P_{\textrm{ND}}\sim e^{-2\pi \Gamma}$ where $\Gamma=(\Delta E)^{2}/4(\textrm{d}\Delta E/\textrm{d}t)$, where $\Delta E$ is the gap of the two levels involved in the crossing. This estimation provides a relation between the rate of variation of the coupling parameter in a protocol, $\textrm{d}g/\textrm{d}t$, and a definite value of the probability $P_{\textrm{ND}}$, namely
\begin{equation}\label{eq:landauzener}
    \frac{\textrm{d}g}{\textrm{d}t}=-\frac{\pi (\Delta E)^{2}}{2\ln P_{\textrm{ND}}}\left(\frac{\textrm{d}\Delta E}{\textrm{d}g}\right)^{-1}.
\end{equation}
For an adiabatic evolution in which $\langle \hatmath{C}\rangle$ changes at each crossing, $P_{\textrm{ND}}\ll 1$.
In our case, a typical crossing shows a gap that changes with the coupling parameter roughly as $\Delta E(g)\sim \omega_{0}g$ [cf. Fig. \ref{fig:requantized}] and therefore for a fixed value of $P_{\textrm{ND}}$, Eq. \eqref{eq:landauzener} implies $\textrm{d}g/\textrm{d}t\sim (\Delta E)^{2}/\omega_{0}$. 
For this reason, an analysis of how the gap $\Delta E$ at an avoided crossing varies with $\omega_{0}$ is required to estimate the value of $|\textrm{d}g/\textrm{d}t|$ required to keep $P_{\textrm{ND}}$ below a certain threshold. 

 To perform these calculations, standard precision algorithms (double precision arithmetic) may be insufficient because the distance of the levels at the avoided crossing can be below their precision limit; as a consequence, higher precision computations are used, which are considerably time-consuming.  We consider two pairs of energy levels close to a given energy at some initial value of the control parameter, $g_{i}\approx 1.9g_{*}$, $\epsilon_{n}(g_{i})$ and $\epsilon_{n+1}(g_{i})$, before the avoided crossing occurs, up to some final value, $g_{f}\approx 2.0g_{*}$, after the crossing has taken place (the precise values depend on $\omega_{0}$). Then, we divide the total span in $g$ into 20 equal parts, and calculate the distance of the two eigenlevels, $|\epsilon_{n}(g_{k})-\epsilon_{n+1}(g_{k})|$, at each of these points. We keep only the smallest of these distances in absolute value, $\Delta E$, and the corresponding value of $g$, $g_m$. Then, we consider a narrower $g$-span between $g_{m-1}$ and $g_{m+1}$, and divide it again into 20 equal parts, to repeat the exact same procedure.
 This is looped for several iterations, and in each iteration we zoom in on the region where the avoided crossing is expected to occur. For an avoided crossing, the distance between eigenlevels $\Delta E$ must saturate to a finite value, as the two levels do not exactly overlap. The results for $\Delta E$, as a function of the resolution $\Delta g=g_{k+1}-g_{k-1}$, are shown in Fig. \ref{fig:deltaEcrossings}(a,c) for different values of $\omega_{0}$; in (a), the pair of levels studied are close to $\epsilon=-4$ for all $\omega_{0}$, while in (c) they are close to $\epsilon=-2$. The extremely small values where the saturation of $\Delta E$ occurs in (a) are remarkable; for $\omega_{0}=10$, the distance of the pair of levels at the crossing is $\Delta E\sim 10^{-13}$, which further decreases as $\omega_{0}$ is increased, and to resolve this avoided crossing one needs to consider the evolution of the levels within a width of $\Delta g\sim 10^{-14}$, which is already a very small variation of the coupling parameter. We emphasize that for $\omega_{0}=15$, which is relatively far away from the thermodynamic limit, the saturation distance is already below the standard numerical precision limit, $\Delta E\sim 10^{-17}$. In (c), $\Delta E$ shows the same qualitative behavior as in (a), but the gap between levels is larger. The value at which $\Delta E$ saturates will be now denoted $\Delta E_{\times}$; this value estimates the gap of the pair of levels at the avoided crossing. This is represented in Fig. \ref{fig:deltaEcrossings}(b,d) as a function of $\omega_{0}$, directly obtained from panels (a,c), respectively. In both cases, this level gap exhibits an exponential decay of the form $\Delta E_{\times}\sim 10^{-\delta \omega_{0}}$, $\delta>0$, indicating that the \textit{avoided crossings are transformed exponentially into real crossings} as $\omega_{0}$ increases, $\Delta E_{x}\to 0$. The value of $\delta$ depends on the energy around which the avoided crossing takes place; $\delta$ decreases as the logarithmic ESQPT around $\epsilon\approx -1$ is approached, as the ESQPT is only fully realized in the limit $\omega_{0}\to\infty$.

 After this analysis, we may estimate the rate of variation of $g$ for an adiabatic process, that is, for a process in which the wavefunction does not jump at every avoided crossing to conserve the value of $\langle \hat{\mathcal{C}}\rangle $. From Eq. \eqref{eq:landauzener}, we have $\textrm{d}g/\textrm{d}t\sim (\Delta E)^{2}/\omega_{0}\sim 10^{-2\delta \omega_{0}}/\omega_{0}$, which vanishes exponentially in $\omega_{0}$ and remains below the standard numerical precision even for small values of $\omega_{0}$. This means that for real processes $P_{\textrm{ND}} \sim 1$ at each avoided crossing, and therefore the conservation of $\langle \hat{\mathcal{C}}\rangle$ is almost perfect. Furthermore, as the number of avoided crossings for a fixed protocol $g_{\textrm{ini}} \rightarrow g_{\textrm{fin}}$ increases linearly with $\omega_0$, we expect that the conservation of $\langle \hat{\mathcal{C}}\rangle$ becomes exponentially better as $\omega_0$ is increased. In Sec. \ref{sec:generationcatstates} we will perform a slow quench protocol with $\omega_{0}=100$, which is moderate but still away from the thermodynamic limit. In this case, for a probability $P_{\textrm{ND}}=0.5$ Eq. \eqref{eq:landauzener} gives $\textrm{d}g/\textrm{d}t\approx 10^{-70}$ for the crossing at $\epsilon=-2$ and $\textrm{d}g/\textrm{d}t\approx 10^{-240}$ for the crossing at $\epsilon=-4$. These rates are out of reach for an experimental setting and also for numerical simulations.

These results provide an important support to the conjecture we have formulated at the end of Sec. \ref{sec:leveldynamics}. Even in finite-size systems, non-adiabatic transitions between levels with different values of $\langle  \hatmath{C} \rangle$ are highly suppressed. Therefore, we can expect the following behavior for an adiabatic passage through the ESQPTs of the deformed Rabi model. Let us suppose that we start from an initial state narrow in energy with $\epsilon>\epsilon_{c2}$. Then, if we slowly change the coupling constant and cross the first ESQPT, levels having both $\langle  \hatmath{C} \rangle=1$ and $\langle  \hatmath{C} \rangle=-1$ become populated, since $\langle  \hatmath{C} \rangle$ abruptly changes from $\langle  \hatmath{C} \rangle \sim 0$ to $\langle  \hatmath{C} \rangle \sim \pm 1$ at the critical line. Afterwards, due to the conservation of $\langle  \hatmath{C} \rangle$, the wavefunction is split into two independent parts, one corresponding to $\langle  \hatmath{C} \rangle=1$ and another corresponding to $\langle  \hatmath{C} \rangle=-1$, whose respective energies change with $g$ with a different rapidity. This suggests that we can use the unitary time evolution to engineer an energy cat state. This is a task that we undertake in the next section. 

\section{Generation of energy cat states}\label{sec:generationcatstates}

From the results in the previous section, we implement the following protocol. We choose an initial state as the ground-state of the quantum Hamiltonian Eq. \eqref{eq:model}, i.e., $\ket{\Psi(g_{\textrm{ini}})}=\ket{\epsilon_{\textrm{GS}}(g_{\textrm{ini}})}$.  
We choose the initial value of the coupling strength $g_{\textrm{ini}}=2.5g_{*}$. Note that this ground state is non-degenerate, so the wavefunction is approximately well located around the corresponding value of $\langle \hat{q} \rangle$.  We then perform a sudden change of the coupling strength $g_{\textrm{ini}}\to g_{\textrm{fin}}$, called a \textit{quench}, to a final value $g_{\textrm{fin}}=1.05g_{*}(\alpha)$. This quench leads the wavefunction to a region above $\epsilon_{c2}$, so it is reasonable to expect that it is still well located around the corresponding value of $\langle \hat{q} \rangle$. Then, we simulate slow driving $g(t)$ by performing successive quenches of the form $g_{i}\to g_{i+1}$, $i=1,2,...$, such that $g_{1}\equiv g_{\textrm{fin}}$, according to $g_{i+1}=g_{i}+\Delta_{g}$ with a small $\Delta_{g}=2\times10^{-5}$ to suppress non-adiabatic transitions throughout the process. After each quench, the non-equilibrium state is allowed to relax during a time $\tau=10^{6}$ and in general will read ($\hbar\equiv 1$)
\begin{equation}
    \ket{\Psi(g)}=\sum_{n}c_{n}(g)e^{-iE_{n}(g)\tau}\ket{E_{n}(g)},
\end{equation}
where all energies and eigenstates are now those corresponding to a Hamiltonian with a given $g$, not necessarily the initial value $g_{\textrm{ini}}$, that is, $\hat{\mathcal{H}}(g)\ket{E(g)}=E(g)\ket{E(g)}$, and $c_{n}(g_{i})\equiv \bra{E_{n}(g_{i})}\ket{\Psi(g_{i-1})}$. This procedure is less computational demanding than a true time-dependent driving $g(t)$.  

At each value of $g$, one may calculate the population of energy levels obtained as the overlap 
\begin{equation}\label{eq:population}\begin{split}
    P(\epsilon(g))&\equiv\sum_{n}|\bra{\epsilon_{n}(g)}\ket{\Psi(g)}|^{2}\delta(\epsilon-\epsilon_{n})\\&=\sum_{n}|c_{n}(g)|^{2}\delta(\epsilon-\epsilon_{n}).
    \end{split}
\end{equation}
Clearly, for the initial state chosen to be the ground-state of the initial Hamiltonian, one has a peak distribution at the ground-state energy, $P(\epsilon(g=g_{\textrm{ini}}))=\delta(\epsilon-\epsilon_{\textrm{GS}}(g_{\textrm{ini}}))$ with $\epsilon_{\textrm{GS}}(g_{\textrm{ini}})=-13.4128$ (not shown). As the initial state is quenched abruptly to $g_{\textrm{fin}}$ this distribution widens. In Fig. \ref{fig:flow} we have represented with a full yellow line the average energy for the trajectory followed by this initial state as it is quenched for different values of $g$, i.e., the energy $\epsilon_{\textrm{q}}(g)=\bra{\Psi(g)}\hat{\mathcal{H}}(g)\ket{\Psi(g)}/(\omega_{0}j)$. This quantity decreases as $g$ increases as a consequence of the curvature of the level flow diagram. The energy distribution $P(\epsilon(g))$ can be calculated for each value of $g$. 

\begin{center}
\begin{figure}[h!]
\hspace*{-0.60cm}\includegraphics[width=0.54\textwidth]{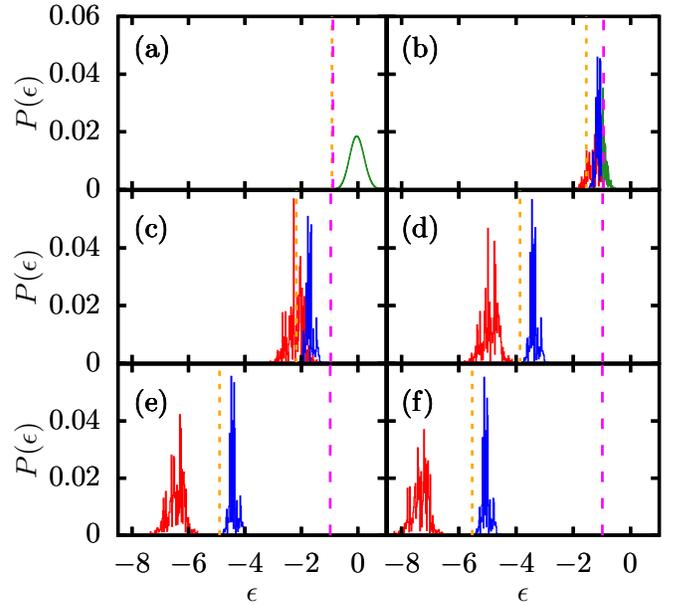}
\caption{Distribution of populated energy levels by the quenched state $\ket{\Psi(g)}$ at a certain value of the coupling strength, Eq. \eqref{eq:population}. Model parameters are $\omega=1$, $\omega_{0}=100$, and $\alpha=1/2$. The value of the coupling strength $g/g_{*}$ is (a) 1.05, (b) 1.34, (c) 1.54, (d) 1.94, (e) 2.14, and (f) 2.25.  
Orange (dotted) and magenta (dashed) vertical lines mark the ESQPTs critical energies $\epsilon_{c1}$ and $\epsilon_{c2}$ for each case [cf. Fig. \ref{energias}]. Number of photons for the whole process is $1900$.}
\label{fig:distributions}
\end{figure}
\end{center}

In Fig. \ref{fig:distributions} we show six cases of the full distribution, schematically indicated in Fig. \ref{fig:flow} with green points defined in the plane $g/g_{*}\times\epsilon_{q}(g/g_{*})$. To facilitate the reading of plots, we have represented with vertical lines two special energy values: the ESQPTs critical energies $\epsilon_{c1}$ (orange [dotted]) and $\epsilon_{c2}$ (magenta [dashed]). 
Green (medium gray) distribution indicates that the populated eigenstates have not been assigned any conserved quantum number by $\hat{\mathcal{C}}$ because it is not a constant of motion at that value of the energy; red (light gray) and blue (dark gray) distributions indicate that the corresponding eigenstates have $\langle\hat{\mathcal{C}}\rangle\leq -0.95$ and $\langle\hat{\mathcal{C}}\rangle\geq 0.95$, respectively. The chosen values of $g$ increase from panel (a) to panel (f) (see caption for details). Panel (a) depicts $P(\epsilon)$ after the first quench $g_{\textrm{ini}}\to g_{\textrm{fin}}$. We find that the distribution of populated energy levels closely resembles a Gaussian distribution with mean $\langle \epsilon\rangle\approx 0$. The form of the distribution stems from the fact that the initial state at $g_{\textrm{ini}}$ is the ground-state of the system which can be seen as a coherent state. This distribution has already widened in panel (b) as a consequence of non-adiabatic transitions, and it has almost completely crossed the logarithmic ESQPT at $\epsilon_{c2}$. For $\epsilon\leq \epsilon_{c2}$ the classical phase space is already composed of separate energy wells and therefore a given eigenstate must necessarily belong to either one, as indicated by $\langle\hat{\mathcal{C}}\rangle$. For this reason, we observe that two separate modes of the distribution $P(\epsilon)$ (in red [light gray] vs. blue [dark gray]) are starting to show up. In panel (c) the initial distribution has completely crossed the logarithmic ESQPT at $\epsilon_{c2}$. We notice that the part of $P(\epsilon)$ corresponding to $\langle\hat{\mathcal{C}}\rangle=-1$ has lower average energy than the part with $\langle\hat{\mathcal{C}}\rangle=+1$. As explained in the previous section, this is because the first set of eigenstates have eigenlevels decreasing faster with $g$ than the second set, i.e., the level flow diagram displays two distinct level dynamics with different curvatures as Fig. \ref{fig:flow} illustrates. As $g$ is further increased, the separation between the two distinct modes of the distribution $P(\epsilon)$ increases. In particular, in panel (d) we observe that the bimodal structure of $P(\epsilon)$ is apparent. Importantly, it is clearly shown how the part of $P(\epsilon)$ with $\langle\hat{\mathcal{C}}\rangle=-1$ has \textit{crossed} the first critical line at $\epsilon_{c1}$, while the part of $P(\epsilon)$ with $\langle\hat{\mathcal{C}}\rangle=+1$ gets trapped before this barrier. This is because when the package gets close to the ESQPT at $\epsilon_{c1}$, only the mode of the distribution of populated levels with $\langle \hat{\mathcal{C}}\rangle=-1$ will be able to pass through, and the mode with $\langle \hat{\mathcal{C}}\rangle=+1$ will be inevitably restrained above this critical energy, as energy levels with $\langle\hat{\mathcal{C}}\rangle=+1$ are not allowed below $\epsilon_{c1}$ [cf. Fig. \ref{fig:flow}]. The result is that as time goes by the two modes of a new bimodal distribution $P(\epsilon)=P(\epsilon_{+})+P(\epsilon_{-})$ will be increasingly further apart in energy. This picture is confirmed in panels (d)-(f), which show that the red (light gray) mode of $P(\epsilon)$ decreases as $g$ increases with the only restriction that $P(\epsilon)=0$ if $\epsilon<\epsilon_{\textrm{GS}}$, but the blue (dark gray) one remains trapped before the barrier located at $\epsilon_{c1}$. The result is the formation of  energy cat states. 

\begin{center}
\begin{figure}[h!]
\hspace*{-0.60cm}\includegraphics[width=0.50\textwidth]{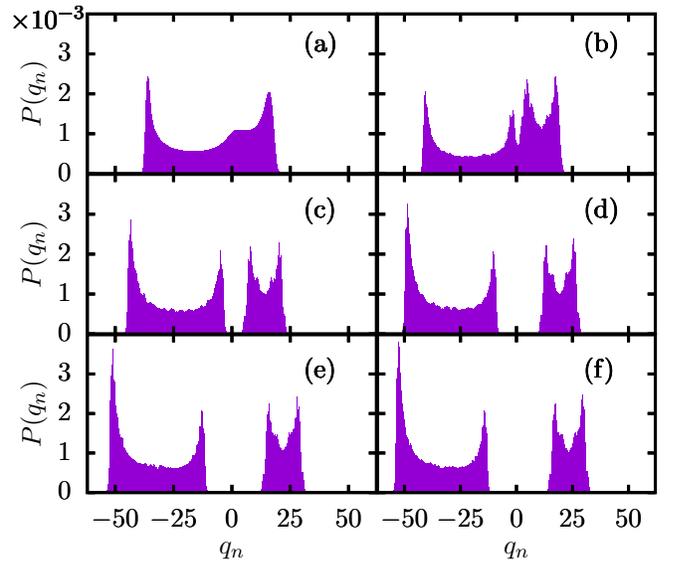}
\caption{Probability Eq. \eqref{eq:probqn} of finding the quenched state $\ket{\Psi(g)}$ at position $q_{n}$. Model parameters are $\omega=1$, $\omega_{0}=100$, and $\alpha=1/2$. The values of the coupling strength $g/g_{*}$ are the same as in Fig. \ref{fig:distributions}: (a) 1.05, (b) 1.34, (c) 1.54, (d) 1.94, (e) 2.14, and (f) 2.25. Number of photons is $1900$. }
\label{fig:probQ}
\end{figure}
\end{center}

To end this section, we link the formation of energy cat states with the semiclassical limit. To do so, we make use of the bosonic quadrature $\hat{q}=(\hat{a}^{\dagger}+\hat{a})/\sqrt{2}$ and solve the eigensystem $\hat{q}\ket{q_{n}}=q_{n}\ket{q_{n}}$. Note that, unlike the classical coordinate $q$, the spectrum of $\hat{q}$ is discrete, i.e., the eigenvalues $q_{n}$ do not form a continuum. The eigenvalues $q_{n}$ are related to the position of the wavefunction in the classical two-dimensional phase space. Using the same quench protocol as before, we calculate the time-averaged probability that the quenched wavefunction $\ket{\Psi(g)}$ be found at $q_{n}$, i.e.,

\begin{equation}\label{eq:probqn}P(q_{n})=\sum_{m}P(\epsilon_{m})|\bra{q_{n}}\ket{\epsilon_{m}}|^{2}\end{equation}
at each step of the quench protocol (i.e., at a given $g$). The probability $P(q_{n})$ is shown in Fig. \ref{fig:probQ} for several cases. These results are analogous to those in Fig. \ref{fig:distributions}, except they are connected with a probability in the classical phase space. The probability Fig. \ref{fig:probQ}(a) shows that the wavefunction can explore both regions of the classical phase space. This is because at this energy the wavefunction is above $\epsilon_{c2}$, and thus classical contour curves consist of two connected wells [cf. Fig. \ref{contours}]. In Fig. \ref{fig:probQ}(b) a visible but still small dip has been formed near $q_{n}=0$. This indicates that the wavepacket is starting to separate, and since this is a probability connected to the classical phase space, this points at a \textit{spatial} separation of the packet. In the rest of panels of Fig. \ref{fig:probQ}(c-f) we observe that the separation of the modes increases as $g$ increases. The reason is that, once the ESQPT at $\epsilon_{c2}$ has been crossed, the wavefunction splits into wavepackets with a definite value of $\langle \hat{\mathcal{C}}\rangle$ as we showed in Fig. \ref{fig:distributions} for $P(E_{n})$. Since the double well structure of the classical phase space is asymmetric when $\alpha\neq 0$, we find that the probabilities $P(q_{n})$ for $q_{n}$ at each side of the phase space are different; in particular, $P(q_{n})$ is wider for $q_{n}<q_{c2}$. This shows that the energy cat states induced by the parity-breaking ESQPT entails the formation of spatial cat states as well, i.e., a coherent superposition of two macroscopically distinct states. 

\section{Thermodynamics of energy cat states}\label{sec:thermodynamics}

The previous results on the emergence of energy cat states generated by the level crossings induced by the ESQPT at $\epsilon_{c2}$ and the constancy of $\hat{\mathcal{C}}$ hint towards an unusual mechanism for thermalization of physically relevant observables, which we study here. 

Generically, quantum thermalization refers to the process by which the expectation value of a physical observable attains a stable equilibrium value around which it simply oscillates for sufficiently long times \cite{Alessio2016}; this equilibrium value coincides with its long-time average. For an observable $\hat{\mathcal{O}}$ in a time-evolving wavefunction $\ket{\Psi(t)}=e^{-i\hat{\mathcal{H}}t}\ket{\Psi(t=0)}=\sum_{n}c_{n}e^{-i E_{n}t}\ket{E_{n}}$ ($\hbar\equiv 1$), this long-time average can be cast in the form
\begin{equation}\label{eq:longtime}
\begin{split}
    &\overline{\langle \hat{\mathcal{O}}\rangle}=\lim_{\tau\to\infty}\frac{1}{\tau}\int_{0}^{\tau}\textrm{d}t\,\bra{\Psi(t)}\hat{\mathcal{O}}\ket{\Psi(t)}\\&=\sum_{n}P(E_{n})\bra{E_{n}}\hat{\mathcal{O}}\ket{E_{n}}
\end{split}\end{equation}
where $P(E_{n})\equiv |c_{n}|^{2}$ is the distribution of populated energy levels, with $c_{n}=\bra{E_{n}}\ket{\Psi(0)}$. In writing the second line of Eq. \eqref{eq:longtime} it has been assumed that the Hamiltonian has no level degeneracies for simplicity, and also because this is the case for our system. Equation \eqref{eq:longtime} implies that the long-time average is determined \textit{entirely} by the distribution $P(E_{n})$ and the diagonal expectation values $O_{nn}\equiv \bra{E_{n}}\hat{\mathcal{O}}\ket{E_{n}}$. Naturally, $\overline{\langle \hat{\mathcal{O}}\rangle}$ depends strongly on those values $O_{nn}$ for which the distribution $P(E_{n})$ is higher, while the expectation values at energies with little population $P(E_{n})\approx 0$ will contribute less to the full average. Once the diagonal expectation values $O_{nn}$ have been fixed, the dynamics is fully governed by the distribution of populated states. This is the essence of the so-called \textit{diagonal ensemble}: one may calculate the exact long-time average of an observable without actually considering the time-evolution of the wavefunction, focusing only on the probability that the wavefunction populates a given eigenstate of the Hamiltonian at any given time, $P(E_{n})$, as well as the diagonal elements of the observable under consideration. This equivalence is well founded mathematically as long as the system exhibits no degeneracies \cite{Alessio2016}

\begin{center}
\begin{figure}[h!]
\hspace*{-0.60cm}\includegraphics[width=0.54\textwidth]{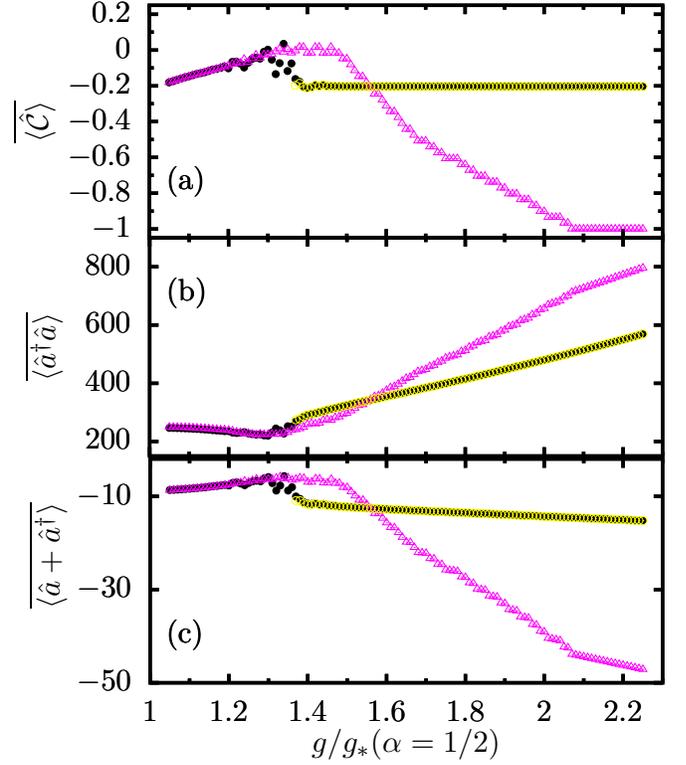}
\caption{Thermalization of physically relevant observables. Full black points represent the exact long-time average Eq. \eqref{eq:longtime}; empty triangles represent the standard microcanonical ensemble Eq. \eqref{eq:micro}; and empty circles represent the result for the modified microcanonical ensemble Eq. \eqref{eq:micromod}. Model parameters are $\omega=1$, $\omega_{0}=100$, and $\alpha=1/2$. The microcanonical ensemble has been obtained by considering a window of $30$ energy levels to each side of the mean energy ($N=61$). In the modified microcanonical ensemble each window has size $15$ ($N_{+}=N_{-}=31$); $p_{+}=0.3984$ and $p_{-}=0.6016$. Number of photons is $1900$. }
\label{fig:paneltermo}
\end{figure}
\end{center}

In a large class of quantum systems, the long-time average Eq. \eqref{eq:longtime} is known to converge to an average obtained from a standard \textit{microcanonical ensemble} centered at a given energy,
\begin{equation}\label{eq:micro}
    \overline{\langle \hat{\mathcal{O}}\rangle}_{\textrm{ME}}=\frac{1}{N}\sum_{E_{n}\in[E-\Delta E,E+\Delta E]}\bra{E_{n}}\hat{\mathcal{O}}\ket{E_{n}},
\end{equation}
where $E$ represents the mean macroscopic energy $\sum_{n}P(E_{n})E_{n}$, and $1\ll N<\infty$ is the number of states contained in a window centered at $E$, $[E-\Delta E,E+\Delta E]$, of width $\Delta E/E\ll 1$. This is certainly the case, for example, in generic quantum chaotic systems, where the equivalence between Eqs. \eqref{eq:longtime} and \eqref{eq:micro} is sustained by the so-called eigenstate thermalization hypothesis \cite{Alessio2016}, and one can also expect the equivalence to hold true for generic, well-behaved distributions $P(E_{n})$ (for example, with a Gaussian form) for which the mean energy is a statistically significant quantity. Yet, there are cases where the simple microcanonical ensemble Eq. \eqref{eq:micro} fails to describe the long-time average Eq. \eqref{eq:longtime}, such as in systems displaying an extensive number of conserved quantities. To treat these systems other statistical ensembles have been devised such as, e.g., the generalized Gibbs ensemble  \cite{Rigol2007} or the generalized microcanonical ensemble \cite{Cassidy2011}. 

Here we focus on the consequences of energy cat states for thermalization. To do so, we study how the expectation value of different physical observables evolves as the slow quench protocol discussed in the previous section is performed. At each step of this successive quench protocol, the initial state is taken to be the final state of the previous step of the protocol, and the corresponding probability distribution of the eigenstates of the final Hamiltonian, $P(E_{n})$, is calculated at each step. Then, use of Eqs. \eqref{eq:longtime} and \eqref{eq:micro} is made to calculate the exact long-time averages and the microcanonical value; in particular, the diagonal ensemble is used to obtain the long-time averages. In Fig. \ref{fig:paneltermo} we compare the exact long-time averages with the standard microcanonical ensemble and with a generalization devised for dealing with energy cat states (see below for details). We focus first on Fig. \ref{fig:paneltermo}(a) which shows the behavior of the operator $\hat{\mathcal{C}}$. It constitutes a remarkable result. We can see that the value of $\langle \hatmath{C} \rangle$ gets stuck at $\langle \hatmath{C} \rangle \sim -0.2$ below $\epsilon_{c2}$ (black points) due to the conservation of $\hatmath{C}$. In the same panel, the standard microcanonical average is also shown (empty triangles). First of all, we can see that this average provides a very poor description of the actual results, but this was somehow expected. So, let us focus on what happens for $g/g_* \gtrsim 2.1$. There, the microcanonical average for $\langle \hatmath{C} \rangle$ reaches its minimum possible value, $\langle \hatmath{C} \rangle=-1$. This means that all the energy levels, $\epsilon_n$, within a small window around the expected energy, $\langle E \rangle$, have $\bra{\epsilon_{n}}\hatmath{C}\ket{\epsilon_{n}}=-1$. Therefore, {\em it becomes impossible to build a generalized ensemble that properly accounts for the actual value of the constant of motion, $\langle \hatmath{C} \rangle \sim -0.2$, just by weighting the population of each energy level within a small energy window around $\langle E \rangle$ according to its value of $\langle \hatmath{C} \rangle$; such a procedure would always yield $\langle \hatmath{C} \rangle=-1$}. Note that this weighting is precisely the method used to build the generalized microcanonical ensemble in \cite{Cassidy2011}, and it is also the basis for the generalized Gibbs ensemble\footnote{At least under the conditions where microcanonical and canonical ensemble become equivalent, the generalized Gibbs ensemble gives rise to a distribution which is only significantly populated within a small energy window around the expected value for the energy, with an irregular shape determined by the expected values of other constants of motion.}. Therefore, a different statistical ensemble is required to deal with energy cat states. 

The reason behind this anomalous behavior is the bimodal structure of the distribution $P(\epsilon_n)$ given by the energy cat state. So, to derive a new kind of generalized microcanonical ensemble describing the thermodynamics of energy cat states, let us consider that the energy distribution is split $P(E_{n})=P(E_{n,+})+P(E_{n,-})$ where $E_{n,\pm}$ denotes the states with $\langle \hat{\mathcal{C}}\rangle=\pm 1$, respectively. Thus, $\mathcal{P}(E_{n,\pm})\equiv P(E_{n,\pm})/\sum_{n}P(E_{n,\pm})$ are the corresponding probability distributions. The difference with $P(E_{n})$ is that each $\mathcal{P}(E_{n,\pm})$ is a unimodal distribution well centered about its mean (in particular, closer to a Gaussian distribution) as can be seen from Fig. \ref{fig:distributions}(d)-(f). This means that one may calculate the mean energy of the states with definite charge $\langle \hat{\mathcal{C}}\rangle=\pm$, $\langle E_{\pm}\rangle=\sum_{n}E_{n,\pm}\mathcal{P}(E_{n,\pm})$ which, unlike $\langle E \rangle$ for $P(E_{n})$, is a good statistical measure once the cat states have been properly formed. This becomes equivalent to performing \textit{two simultaneous} microcanonical averages, each one centered about each of the two means $\langle E_{n,+}\rangle$ and $\langle E_{n,-}\rangle$, and normalize the result correspondingly. This is realized in the form of the following modified microcanonical ensemble:
\begin{equation}
\label{eq:micromod}
\begin{split}
\overline{\langle \hat{\mathcal{O}}\rangle}_{\textrm{ME2}}&=\frac{p_{+}}{N_{+}}\sum_{E_{n,+}\in[E_{+}-\Delta E_{+},E_{+}+\Delta E_{+}]}\bra{E_{n,+}}\hat{\mathcal{O}}\ket{E_{n,+}}+ \\ &+\frac{p_{-}}{N_{-}}\sum_{E_{n,-}\in[E_{-}-\Delta E_{-},E_{-}+\Delta E_{-}]}\bra{E_{n,-}}\hat{\mathcal{O}}\ket{E_{n,-}},
\end{split}
\end{equation}
with
\begin{equation}
p_{\pm}\equiv \frac{1\pm\overline{\langle \hat{\mathcal{C}}\rangle}}{2}.
\end{equation}
The normalization factors $p_{\pm}$ are the probability that a given wavefunction be fully localized within the left $(-1)$ or right ($+1$) classical energy well. In general, a wavefunction will be a superposition of those two limiting cases since $-1\leq \bra{\Psi}\hat{\mathcal{C}}\ket{\Psi}\leq 1$, and thus it follows that $0\leq p_{\pm}\leq 1$ and $p_{+}+p_{-}=1$. The rest of Eq. \eqref{eq:micromod} has the same interpretation as in the standard microcanonical ensemble Eq. \eqref{eq:micro}, with the exception that now two averages are performed instead of just one, within two different energy windows (possibly) containing different number of levels. One caveat of the ensemble Eq. \eqref{eq:micromod} is that it heavily depends on the separation in energy subspaces allowed by $\hat{\mathcal{C}}$, and thus it is \textit{undefined} where $\hat{\mathcal{C}}$ is not a constant of motion. Also note that, by construction, $\overline{\langle \hat{\mathcal{C}}}\rangle_{\textrm{ME2}}=p_{+}-p_{-}=\overline{\langle\hat{\mathcal{C}}\rangle}$.

Now, we come back to Fig. \ref{fig:paneltermo}. Besides the expected value of $\langle \hatmath{C} \rangle$, we show there $\langle a^{\dagger} a \rangle$, in panel (b), and $\langle a^{\dagger} + a \rangle$, in panel (c). The generalized microcanonical ensemble, provided by Eq. \eqref{eq:micromod}, is represented by empty yellow circles, and has been calculated only once $\hat{\mathcal{C}}$ acts a constant of motion; for $g/g_{*}(\alpha=1/2)\lesssim 1.4$, it is undefined [cf. Fig. \ref{fig:flow}]. We can see that for small values of $g$ close to $g_{\textrm{fin}}=1.05g_{*}(\alpha=1/2)$, the long-time and the standard microcanonical averages agree very well. However, as $g$ increases the microcanonical value starts showing significant deviations with the exact long-time average and for values $g/g_{*}(\alpha=1/2)\gtrsim 1.3$ the previous accordance is totally ruined. On the contrary, the generalized microcanonical ensemble given by Eq. \eqref{eq:micromod} provides a perfect description for all the three studied observables when $\hat{\mathcal{C}}$ acts like a constant of motion.

Therefore, we conclude that in order to describe the thermalization of physical observables in the energy cat states described in this work, a generalized version of the microcanonical ensemble, Eq. \eqref{eq:micromod}, characterized by two different expected values for the energy, corresponding to the two parts of the bimodal structure of the energy distribution, has to be used instead of the common Eq. \eqref{eq:micro}. This implies that the thermodynamics of energy cat states is characterized by two different internal energies, which give rise to two different temperatures. For a standard isolated system with energy $E$, we can define a microcanonical temperature \cite{Pathria}, $1/T=\partial S (E)/\partial E$, where $S(E)$ is the entropy, and the Boltzmann constant is set to $k_{B}\equiv 1$. In our case, as we need two different energies to build thermodynamics, $E_+$ and $E_-$, the entropy must be a function of these two energies, $S(E_+,E_-)=\ln \left[ \rho_+(E_+) + \rho_-(E_-) \right]$, where $\rho_{\pm} (E_{\pm})$ are the parts of the density of states corresponding to energy levels with $\langle \hat{\mathcal{C}}\rangle=\pm 1$. Hence, there exist two different temperatures, $1/T_{\pm}=\partial S (E_+,E_-)/\partial E_{\pm}$, each one evaluated at the corresponding value of the average energy of the bimodal distribution of the cat state.

\section{Conclusions}\label{sec:conclusions}

The main result of this work is a protocol to create an energy cat state ---a Schr\"odinger cat state involving a quantum superposition of both different positions and energies--- by slowly crossing two different ESQPTs. To do so, we rely in a generalization of the Rabi model which includes a parity-breaking term. As a point of departure, we have shown that this model has two different ESQPTs, at energies $\epsilon_{c2} \geq \epsilon_{c1}$, without any kind of QPT. The corresponding phases can be described by means an observable with just two eiganvalues, $\hatmath{C}$ which is a constant of motion below $\epsilon_{c2}$. In the phase with $\epsilon_{c1} \leq \epsilon \leq \epsilon_{c2}$, level crossings between energy levels with $\langle \hatmath{C} \rangle = 1$ and $\langle \hatmath{C}\rangle =-1$ are observed in the thermodynamic limit, when the coupling constant $g$ is changed. In the phase with $\epsilon<\epsilon_{c1}$ all the energy levels have $\langle \hatmath{C}\rangle = -1$. 

By means of stringent numerical calculations we have shown how to engineer an energy cat profiting from the previous physical situation. First, the protocol is started in the ground state of the generalized Rabi system with a large value of the coupling constant $g$. Second, a quench onto a smaller value of this constant is performed; as a result, the system equilibrates above the critical energy associated with the logarithmic ESQPT, $\epsilon_{c2}$. Then, the coupling constant is slowly increased. As a consequence, the wavefunction enters the phase with $\epsilon_{c1} \leq \epsilon \leq \epsilon_{c2}$. As the energy of levels with $\langle \hatmath{C}\rangle =-1$ changes faster with $g$ than that of levels with $\langle \hatmath{C}\rangle =1$, the wavefunction splits into two different parts: one centered at $q<q_c$ with lower energy, and another centered at $q>q_c$ and higher energy. Finally, at the final stages of the protocol, only the first part of the wavefunction crosses the critical energy associated with the finite discontinuity in the level density, $\epsilon_{c1}$. As a result, the larger the final value of $g$, the more separated are the two parts of the wavefunction. The result is an energy cat in which two different positions and two different energies are superposed.

Finally, we have studied the thermodynamics of this state. We show that a new kind of ensemble, including two different average energies (one for the part of the wavefunction with $\langle \hatmath{C}\rangle =-1$ and the other for the part with $\langle \hatmath{C}\rangle =1$), is required to properly describe the resulting equilibrium state. 

\begin{acknowledgments}
We gratefully thank F. P\'{e}rez-Bernal and J. Dukelsky for insightful  discussions. This work has been supported by the Spanish Grant No. PGC-2018-094180-B-I00 funded by Ministerio de Ciencia e Innovaci\'{o}n/Agencia Estatal de Investigaci\'{o}n MCIN/AEI/10.13039/501100011033 and FEDER "A Way of Making Europe". A. L. C. acknowledges financial support from `la Caixa' Foundation (ID 100010434) through the fellowship LCF/BQ/DR21/11880024.
\end{acknowledgments}

\end{document}